\documentclass[12pt]{iopart}
\usepackage{graphicx}
\usepackage{dcolumn}
\usepackage{bm}
\usepackage{epstopdf}
\usepackage[dvipsnames]{xcolor}
\usepackage{soul}
\usepackage[utf8]{inputenc}
\usepackage[T1]{fontenc}
\usepackage{indentfirst}
\usepackage{har2nat}
\usepackage{enumitem}
\usepackage{xspace}
\usepackage{hyperref}
\hypersetup{breaklinks=true,colorlinks=true,linkcolor=blue,citecolor=blue,filecolor=magenta,urlcolor=cyan}
\usepackage{datetime}

\newcommand{\RevModPhys}[1]{}
\newcommand{\JPhysG}[1]{}

\newcommand{\newblock}{}

\usepackage{array}
\newcolumntype{L}[1]{>{\raggedright\let\newline\\\arraybackslash\hspace{0pt}}p{#1}}

\newcolumntype{C}[1]{>{\centering\let\newline\\\arraybackslash\hspace{0pt}}p{#1}}
\newcolumntype{R}[1]{>{\raggedleft\let\newline\\\arraybackslash\hspace{0pt}}p{#1}}

\renewcommand{\harvardurl}[1]{\url{#1}}
\renewcommand{\vec}[1]{\mbox{\boldmath $#1$}}

\newcommand{\Sec}{section}
\newcommand{\Fig}{figure}

\begin{document}

\title[Future of Nuclear Fission Theory]{Future of Nuclear Fission Theory}

\author{%
Michael         Bender$          ^{1}$,
R\'emi          Bernard$       ^{2,3}$,
George          Bertsch$         ^{4}$,
Satoshi         Chiba$           ^{5}$,
Jacek           Dobaczewski$ ^{6,7,8}$,
No\"el          Dubray$        ^{3,9}$,
Samuel A.       Giuliani$       ^{10}$,
Kouichi         Hagino$         ^{11}$,
Denis           Lacroix$        ^{12}$,
Zhipan          Li$             ^{13}$,
Piotr           Magierski$      ^{14}$,
Joachim         Maruhn$         ^{15}$,
Witold          Nazarewicz$     ^{16}$,
Junchen         Pei$            ^{17}$,
Sophie          P\'eru$        ^{3,9}$,
Nathalie        Pillet$        ^{3,9}$,
J\o{}rgen       Randrup$        ^{18}$,
David           Regnier$       ^{3,9}$,
Paul-Gerhard    Reinhard$       ^{19}$,
Luis M.         Robledo$     ^{20,21}$,
Wouter          Ryssens$        ^{22}$,
Jhilam          Sadhukhan$   ^{23,24}$,
Guillaume       Scamps$         ^{25}$,
Nicolas         Schunck$        ^{26}$,
C\'edric        Simenel$         ^{2}$,
Janusz          Skalski$        ^{27}$,
Ionel           Stetcu$         ^{28}$,
Paul            Stevenson$      ^{29}$,
Sait            Umar$           ^{30}$,
Marc            Verri{\`e}re$^{26,28}$,
Dario           Vretenar$       ^{31}$,
Micha{\l}       Warda$          ^{32}$,
Sven            {\AA}berg$      ^{33}$
}
\bigskip
\address{$^{1}$  IP2I Lyon, CNRS/IN2P3, Universit{\'e} de Lyon, Universit{\'e} Claude Bernard Lyon 1, F-69622 Villeurbanne, France}
\address{$^{2}$  Department of Theoretical Physics and Department of Nuclear Physics, Research School of Physics, Australian National University, Canberra, Australian Capital Territory 2601, Australia}
\address{$^{3}$  CEA, DAM, DIF, F-91297 Arpajon, France}
\address{$^{4}$  Department of Physics and Institute for Nuclear Theory, University of Washington, Seattle, Washington 98195, USA}
\address{$^{5}$  Tokyo Institute of Technology, 152-8550 Tokyo, Japan}
\address{$^{6}$  Department of Physics, University of York, Heslington, York YO10 5DD, United Kingdom}
\address{$^{7}$  Helsinki Institute of Physics, FI-00014 University of Helsinki, Finland}
\address{$^{8}$  Institute of Theoretical Physics, Faculty of Physics, University of Warsaw, 02-093 Warsaw, Poland}
\address{$^{9}$  Universit{\'e} Paris-Saclay, CEA, Laboratoire Mati{\`e}re en Conditions Extr{\^e}mes, Bruy{\`e}res-le-Ch{\^a}tel, France}
\address{$^{10}$ FRIB Laboratory, Michigan State University, East Lansing, Michigan 48824, USA}
\address{$^{11}$ Department of Physics, Kyoto University, Kyoto 606-8502, Japan}
\address{$^{12}$ Universit\'e Paris-Saclay, CNRS/IN2P3, IJCLab, 91405 Orsay, France}
\address{$^{13}$ School of Physical Science and Technology, Southwest University, Chongqing 400715, China}
\address{$^{14}$ Faculty of Physics, Warsaw University of Technology, 00-662 Warsaw, Poland}
\address{$^{15}$ Institut f\"ur Theoretische Physik, Goethe-Universit\"at, Max-von-Laue-Str. 1, 60438 Frankfurt am Main, Germany}
\address{$^{16}$ Department of Physics and Astronomy and FRIB Laboratory, Michigan State University, East Lansing, Michigan 48824, USA}
\address{$^{17}$ Department of Technical Physics, School of Physics, Peking University, Beijing 100871, China}
\address{$^{18}$ Nuclear Science Division, Lawrence Berkeley National Laboratory, Berkeley, California 94720, USA}
\address{$^{19}$ Institut f{\"u}r Theoretische Physik II, Universit\"at Erlangen-N\"urnberg, 91058 Erlangen, Germany}
\address{$^{20}$ Center for Computational Simulation, Universidad Polit\'ecnica de Madrid, E-28660 Madrid}
\address{$^{21}$ Departamento de F\'\i sica Te\'orica and CIAFF, Universidad Aut\'onoma de Madrid, E-28049 Madrid, Spain}
\address{$^{22}$ Center for Theoretical Physics, Sloane Physics Laboratory, Yale University, New Haven, CT 06520, USA}
\address{$^{23}$ Variable Energy Cyclotron Centre, Kolkata 700064}
\address{$^{24}$ Homi Bhabha National Institute, Mumbai 400094, India}
\address{$^{25}$ Institut d'Astronomie et d'Astrophysique, Universit\'e Libre de Bruxelles, Campus de la Plaine CP 226, BE-1050 Brussels, Belgium}
\address{$^{26}$ Nuclear and Chemical Science Division, Lawrence Livermore National Laboratory, Livermore, California 94551, USA}
\address{$^{27}$ National Center for Nuclear Research, Pasteura 7, 02-093 Warsaw, Poland}
\address{$^{28}$ Theoretical Division, Los Alamos National Laboratory, Los Alamos, NM 87545, USA}
\address{$^{29}$ Department of Physics, University of Surrey, Guildford, Surrey, GU2 7XH, United Kingdom}
\address{$^{30}$ Department of Physics and Astronomy, Vanderbilt University, Nashville, TN 37235, USA}
\address{$^{31}$ Department of Physics, Faculty of Science, University of Zagreb, Croatia}
\address{$^{32}$ Institute of Physics, Maria Curie-Sk{\l}odowska University, 20-031 Lublin, Poland}
\address{$^{33}$ Mathematical Physics, Lund University, S-221 00 Lund, Sweden}


\date{\today{}}

\ead{jacek.dobaczewski@york.ac.uk}

\begin{abstract}
There has been much recent interest in nuclear fission, due in part to a new appreciation of its relevance to astrophysics, stability of superheavy elements, and fundamental theory of neutrino interactions.  At the same time, there have been important developments  on a conceptual and computational level for the theory.  The promising new theoretical avenues were the subject of a workshop held at the University of York in October 2019;  this report summarises its findings and recommendations.

\end{abstract}

\submitto{\JPG}

\maketitle

\tableofcontents{}

\newpage

\markboth{Future of Nuclear Fission Theory}{Future of Nuclear Fission Theory}


\section{Introduction}
\label{sec:intro}

The theory of nuclear fission has a long history, driven
for many years by technological applications and heavy element
studies.  Today the needs are even broader with the recognition
of new connections to other disciplines
such as astrophysics and fundamental science.

In the past, fission theory was largely phenomenological.
Recent significant advances in microscopic modeling, which can be tested
thanks to the rapid growth in computational capabilities including
leadership-class computers, provide opportunities for developing fission theory
to a new level of refinement.
In addition, experimental fission data of unprecedented detail and quality
are now being acquired and can be used to validate models more thoroughly.

A disclaimer  is in order, because  we used the  word ``microscopic''
in the previous paragraph.  In the context of fission theory and
indeed all theory that is
applied to large nuclei, ``microscopic'' should not be construed as an {\it
ab initio} many-body theory with all Hamiltonian input taken from the
outside.  In our field, theory is useful at a quantitative level only if
the parameters, or coupling constants,  of   models are
optimised to  experiment.  For that reason, all quantitative nuclear models are
phenomenological at some level. Superlatives such as `fully
microscopic' or `from first principles', sometimes used to characterise
particular approaches, may be viewed more as wishful thinking than the
present reality.  However, it is useful to distinguish the degrees of
phenomenology   in different theoretical approaches.
In this document we will use the term
``microscopic theory''  for theoretical approaches in which nucleonic
degrees of freedom are explicitly present together with inter-nucleon forces.
The most prominent example is
nuclear density functional theory which is based on
effective nucleon-nucleon interactions that generate mean fields and the
associated  single-particle orbitals.   In this document we assess the future
promise of a number of extensions of density functional theory.  Some of
them remain microscopic, but others are best characterised as
phenomenological.

This document was initiated at the Workshop on {\it Future of
Theory in Fission} held in York in October 2019
(\url{https://www.york.ac.uk/physics/news/events/groups/nuclear-physics/2019/future-of-theory-in-fission-workshop/}).
The premise
of the meeting was that fission theory is ripe for rapid progress.
Consequently, the focus was on future
developments, perspectives, and challenges.
The questions motivating the workshop were:
\begin{itemize}
\item
Considering the broad range of observables, what are the
  physics objectives that fission theory needs to address?
\item
What are realistic goals that can be achieved with advanced microscopic
  frameworks and  modern computational tools?
\item
Can microscopic theory provide justification for successful
  phenomenological assumptions and models?
\item
Which current approximations routinely made in fission
  studies are justified or not, or unavoidable or not, in view of
  the present-day computational capabilities?  What are the
  robust approximations that can be employed to simplify the treatment
  and/or reduce the computational effort?
\item
 Is it realistic to envision a unified microscopic theory of
  fission that would cover the entire energy range
  from spontaneous fission to fission well above the barrier?
\item
What are the best strategies for the community
     to optimise fission theory research?
\end{itemize}

As seen from the table of contents, this document summarises the broad range
of topics covered in the York Workshop.  Our unifying theme is the pathway towards solving the fission problem
via modern many-body frameworks by taking advantage of the latest computational methodologies.
In this context, the main purpose of this document is to outline challenges
and point to possible solutions rather than to provide a detailed review of
nuclear fission theory.
The interested reader is encouraged to consult recent reviews of
various aspects of fission theory, e.g., \citeasnoun{(Sch16b)} (overview of microscopic models) and
\citeasnoun{(And17),Talou2018,Schmidt2018}
(description of state-of-the-art fission phenomenology), which contain extensive
lists of references.

\begin{figure*}[!htb]
\begin{flushright}{\includegraphics[width=0.85\linewidth]{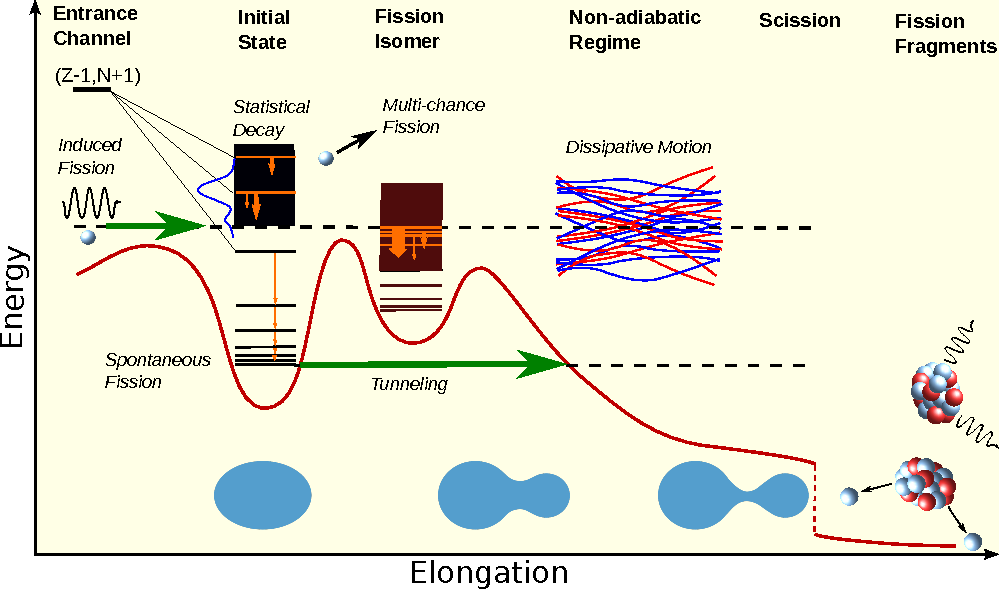}}\end{flushright}
\caption{\label{fig:fission_physics}Schematic illustration
of the features most relevant to the fission phenomenon.
The red curve depicts (in a one-dimensional projection)
the potential energy  as a function of the elongation;
the ground state is at the lowest minimum, and the shape-isomeric state
is at the second minimum.
From these states it is possible to tunnel through the potential barrier.
Tunnelling is also relevant for neutron or photon induced fission
when the resulting initial state lies below the fission barrier.
If the initial state is excited above the fission barrier, it may undergo a complicated
shape evolution crossing the barrier from above.
Once the system finds itself beyond the barrier,
it relatively quickly descends towards scission.
There it divides into two nascent fragments,
which subsequently move apart under the influence of their mutual
Coulomb repulsion while gradually attaining their equilibrium shapes and become primary
fragments. Primary fragments then
de-excite by evaporating neutrons, radiating photons, and undergoing $\beta$ decay.
}
\end{figure*}

Of the various shape regions depicted in {\Fig}~\ref{fig:fission_physics},
we shall cover microscopic
dynamics in the domains around the initial state and the barriers as well as
the highly deformed region beyond.  But we are leaving out the important
challenge of describing how the system propagates from one region to another
because there has been virtually no coherent microscopic theory addressing
this question up to now.  This underscores the fact that there will still
be much future work to do in nuclear fission theory.

\section{Main Features of Fission}
\label{sec:phenomenon}

\begin{figure*}[!htb]
\begin{flushright}{\includegraphics[width=0.85\linewidth]{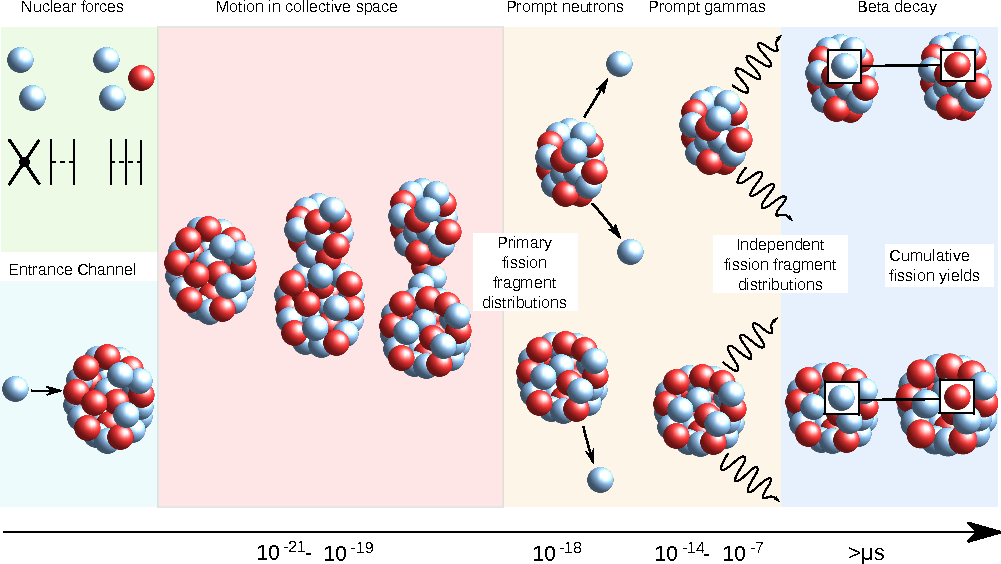}}\end{flushright}
\caption{\label{fig:fission_process}Schematic representation
of the different stages of a fission process,
starting from the initial nucleus (on the left), approaching
the scission point as pre-fragments,
dividing into two excited nascent fragments, which after getting
fully Coulomb accelerated become primary fragments, then
promptly emit neutrons and photons and undergo $\beta$ decays,
and finally become fission fragments in the exit channel.
The associated time scales are indicated on the axis underneath.
}
\end{figure*}

To set the stage for the subsequent specialised considerations,
we begin with a brief presentation of the main features
of the nuclear fission phenomenon.
Figures~\ref{fig:fission_physics} and \ref{fig:fission_process} present
schematic illustrations of the evolution leading from a single nucleus
to two pre-fragments, nascent fragments, primary fragments, which subsequently
appear in detectors as fission fragments, see caption of {\Fig}~\ref{fig:fission_process}.

Fission is a time-dependent
transformation which can be conveniently separated into distinct stages,
each characterised by its own time scale, as shown in
{\Fig}~\ref{fig:fission_process}.
The process proceeds from some initial state
through a complicated collective evolution ending with the emergence
of two excited nascent fragments.  They in turn
undergo a sequence of prompt and/or delayed de-excitations decays
ending with two product nuclei in their ground or isomeric excited
states\footnote{
In this paper, we are not concerned with the extremely rare phenomena of ternary and  quaternary fission.}.

The most obvious physical attribute during the evolution of the fissioning
nucleus
is its overall elongation, correlated with the different stages as shown
in {\Fig}~\ref{fig:fission_physics}.
Initially the elongation is that of the equilibrium shape of the mother nucleus.
From this, the collective evolution proceeds through a sequence of shapes
whose time-dependent elongations exhibit a diffusive behaviour.
Eventually the system finds itself beyond the outer saddle point
and then evolves toward scission,
as its shape takes on a binary form and the elongation grows ever larger.
At scission the system divides into  nascent fragments
which are then accelerated apart.

\subsection{Spontaneous and induced fission}
\label{subsec:sf}
   It is useful to distinguish spontaneous fission (SF) which occurs in nuclei in
their ground states from induced fission brought about by a reaction or
decay process bringing in energy from the outside.
SF is one of the main decay modes of
superheavy nuclei and is therefore of great interest in the experimental
search for them.
While SF primarily occurs from the nuclear ground state,
it has also been observed from isomeric states.

On the theory side, the relatively long lifetimes are due to the existence
of a potential barrier that must be penetrated.  Consequently SF  is
an inherently quantal process; see {\Sec}~\ref{subsec:tunnelling}.  An interesting
aspect of SF is its dependence on the number parity of
the nucleus: in
odd-$A$ nuclei it is typically hindered by $\sim 3 - 5$
orders of magnitude relative to their even-even neighbours.
Fission of  odd-odd nuclei is believed to be even more
hindered, but credible data are scarce.

In addition to a SF, fission can be induced by a variety of nuclear
reactions. The fission-induced processes  include: neutron capture
(responsible for energy production in fission reactors), electron
capture and beta decay, photofission, and reactions involving charged
particles and heavy ions. In all these processes, the fissioning
nucleus is created in an excited state, which may lie above or below
the fission barrier.

Theoretical descriptions of fission induced by fast probes often
assume the creation of a compound nucleus at a given thermal
excitation energy. However, as discussed later, that assumption might
be ill-founded for fast probes because the nuclear system may not
have sufficient time to thermalise before undergoing fission. This
becomes increasingly important at higher energies where
pre-equilibrium processes play an increasingly significant role and
may lead to the emission of one or more nucleons before equilibrium
is reached. Moreover, as the excitation energy of the compound
nucleus is increased, neutron evaporation competes ever more
favourably with fission and as a result, one or more neutrons may be
evaporated before fission occurs (multi-chance fission).
In addition, for non-thermalised systems one should develop
approaches using fixed energy rather than
fixed temperature.

\subsection{Important observables}\label{subsec:observables}

When talking about fission observables, it is important to remember
that what is often considered ``experimental'' is often the result of
an indirect process, in which a quantity of interest is extracted
from measurements with the help of some model or  model-dependent
assumptions.

Nuclear fission is a very complex transformation and there are many
quantities of interest that are directly measurable and subject to theoretical modelling.
[A set of key fission observables
suitable for validation of theoretical models was proposed in \citeasnoun{Bertsch2015}.]
We list here some of the most important ones, with their common
designations:
\begin{description}
\item[Spontaneous-fission half-lives (\boldmath{$T_{\rm SF}$).}] Measured SF lifetimes (or half-lives) span a range from microseconds or smaller to billions of years. To describe such a range is a significant challenge to theory.
\item[Total and differential fission cross sections.] For instance, the neutron
induced fission cross section  $\sigma$(n,f) and its energy and angular  dependence or
the threshold energy for fission observed in a photo-fission cross section
that is closely related to the height of a fission barrier.
\item[Yields (\boldmath{$Y(A)$, $Y(Z)$, $Y(Z,A)$}).] They describe probabilities
      for producing fission fragments of given mass and/or charge.  Such data are
      particularly important in nuclear astrophysics.
      Yields refer to primary, independent or cumulative distributions (see {\Fig}~\ref{fig:fission_process}).
\item[Fission spectrum.]  This includes the average number of neutrons
      per fragment, their energies, the average number of
      photons per fragment and their energies, multiplicity distributions, angular correlations, etc.
\item[Total kinetic energy (TKE).]  The post-acceleration kinetic energy of
      the fission fragments, its distribution, and its dependence on fragment mass.
\item[Beta-decay spectrum of fission products.]  This is particularly
      important for the fundamental theory of beta decay and includes the neutrino spectrum.
\end{description}
Correlations between the above quantities (e.g., between fragment mass and TKE),
as well as with other quantities (e.g., with the spin of the
fissioning nucleus) are also very important.
We wish to emphasise that the fission observables should be accompanied by
uncertainties. This is crucial in the context of nuclear data
evaluation and applications in general.

In this context, it is useful to mention some important
unobservables (physical concepts that cannot be observed directly).
Arguably, the most celebrated quantity that
belongs to this group is the fission barrier. Fission barrier height
can be defined theoretically as the energy difference between the
ground state and the highest saddle point in a computed  potential
energy surface (PES) that has the lowest energy for all possible
paths leading to fission from the ground state. Fission barriers
inferred from measured cross sections   are plagued with ambiguities
because the extraction procedure is often based on  a simplistic
picture of a fission pathway. Another unobservable concept is that of
a  compound nucleus; it is  based on a model that assumes the full
thermalisation of the system  and ignores  pre-equilibrium processes.
Other useful yet unobservable quantities include: scission point at
which the nucleus breaks into nascent fragments, shell energy on the path to
fission, pairing energy at the barrier, and pre-fragments that are
formed in the pre-scission region.


\section{Basic concepts of fission theory}
\label{sec:concepts}

To lay the groundwork for discussing the promising ideas for future
development, we recall here some of the basic theoretical tools at our
disposal.
In time-dependent formalisms, basic distinctions can
be made between dynamics based on inertial motion, dynamics  based on
diffusion motion in a statistical framework, and dynamics  that
combine inertial and diffusive motion.  The relevant computational methodologies are often referred
to by their acronyms; the ones used here are listed in
Table~\ref{Glossary}.
\begin{table}[!htb]
\caption{\label{Glossary} Glossary of used acronyms pertaining to nuclear fission models
and fission characteristics (in alphabetic order).
}
\begin{indented}
\item[]\begin{tabular}{ll}
\br
Acronym &  Meaning \\
\mr
ATDDFT &   Adiabatic TDDFT                          \\
ATDHFB &   Adiabatic TDHFB                          \\
BCS    &   Bardeen-Cooper-Schrieffer                \\
CHF    &   Constrained HF                           \\
CHFB   &   Constrained HFB                          \\
CSE    &   Collective Schr\"odinger equation        \\
DDD    &   Dissipative Diabatic Dynamics            \\
DFT    &   Density Functional Theory                \\
EDF    &   Energy Density Functional                \\
GCM    &   Generator Coordinate Method              \\
GOA    &   Gaussian overlap approximation           \\
HF     &   Hartree-Fock                             \\
HFB    &   Hartree-Fock-Bogoliubov                  \\
HO     &   Harmonic Oscillator                      \\
MM     &   Microscopic Macroscopic                  \\
PES    &   Potential Energy Surface                 \\
QRPA   &   Quasiparticle RPA                        \\
RPA    &   Random Phase Approximation               \\
SF     &   Spontaneous Fission                      \\
TDDFT  &   Time-Dependent DFT                       \\
TDGCM  &   Time-Dependent GCM                       \\
TDHF   &   Time-Dependent HF                        \\
TDHFB  &   Time-Dependent HFB                       \\
TDRPA  &   Time-Dependent RPA                       \\
TKE    &   Total Kinetic Energy                     \\
\br
\end{tabular}
\end{indented}
\end{table}

Before going into details of different concepts discussed below, we
want to touch upon one specific term that is abundantly used in the
theory of nuclear fission, namely, the concept of adiabaticity.
First, a disclaimer is in order, because in the rigorous (electronic)
time-dependent density functional theory (TDDFT), see,
e.g.~\citeasnoun{(Bur05a)}, the term "adiabatic" has a different
meaning than here. Indeed, there it denotes an approximation of the
time-dependent functional that is local in time and thus disregards
memory effects. In this sense, all time-dependent approaches to
fission, which we discuss below, are adiabatic, and releasing this
constraint in nuclear physics probably belongs to the future not
covered by the present report at all.

In nuclear physics, the term "adiabatic" has several interwoven,
although not fully identical facets. First, it may mean that the
collective motion proceeds through a sequence of local ground states,
each corresponding to the system being constrained to a given set of
collective coordinates and intrinsic quantum numbers. Adiabatic
motion then means that a time-dependent wave function acquires
collective kinetic energy through infinitesimal admixtures of local
excited states, whereas non-adiabatic corrections correspond to
significant admixtures of those. A dissipative motion
({\Sec}~\ref{dissdyn}) means a constant irreversible flow of energy
away from the local ground state.

Provided the local ground states
are well defined and do not cross with excited states, this
constitutes a coherent physical picture. However, this picture breaks
down in situations where several local ground states (characterized
by different intrinsic quantum numbers) coexist and compete
energetically (e.g., different one-quasiparticle states in odd-$A$
nuclei). Then, the system may proceed diabatically, along a fixed
configuration, or adiabatically, by changing the configuration,
depending on the  Landau-Zener probability of a diabatic transition
({\Sec}~\ref{Landau-Zener}).

In  the context of the time-dependent Hartree-Fock (TDHF) or
TDDFT, adiabaticity denotes a very specific approximation of the
time-dependent one-body density matrix, which is assumed to have the
time-odd part much smaller than its time-even part \cite{(Bar78b)}.
In essence, this approximation holds only when the motion is
appropriately slow. Another commonly used definition of adiabaticity involves a
separation of variables into slow and fast coordinates
\cite{Tully2012}. Many concepts of fission theory, such as the
Collective Schr{\"o}dinger Equation (CSE), are based on the division
of degrees of freedom into ``collective'' and "non-collective''.

All those definitions are connected by the fact that, in
practice, the local ground states can only be considered within the
mean-field picture, which means the TDDFT interpretation of the
one-body evolution. A weak mixing with low-lying
excited states is then equivalent to the requirement of the slow
motion. In the following, the notions of adiabaticity and
dissipation are discussed in many places, as undoubtedly they
constitute pivotal points of the theoretical description of fission.


\subsection{Time scales}
\label{subsec:times}

It is important to understand the various time scales associated
with the different stages of fission in order to anticipate the
kind of dynamics that would be needed in the theory.
One of the most intriguing questions about fission dynamics is the
time it takes for fission to occur.

There are in fact several time scales that affect the duration of the fission
process.  Fission that goes through the compound nucleus  is delayed by the compound nucleus lifetime, which is
 much longer than the dynamics time scales.
At excitation energies below the fission barrier, the fission
lifetime is largely dominated by the tunnelling probability and can vary
by many orders of magnitude. The next scale is that of the
collective motion from the outer turning point  to  scission,
see {\Fig}~\ref{fig:fission_physics}.  The slower
this is, the more valid will be diffusive and statistical modelling
of the dynamics.  Finally,
the time it takes to scission plays a special role affecting particularly
the TKE and excitation energies of the fragments.
At higher energies, the distinctions between the different stages are
less clear, but the basic dynamics taking the system from a highly
excited compound nucleus to a scission configuration is governed
by a similar time scale.

One of the most difficult questions to investigate experimentally is fission
time scales since they involve
the early stages of fission dynamics. They are not generally accessible
directly but must be inferred from the analysis of products at later
stages of fission.
Experiments attempting to measure fission times \cite{(Hin93),Jacquet2009,(Fre12),Sikdar2018} often need to be complemented
by a model description of, e.g., the emitted neutrons and their dependency on
angular momentum or excitation energy.  See {\Sec}~\ref{subsec:spectrum} for a
discussion of this topic. As a result, it is likely that different experimental methods
probe different characteristics of the fission time distribution.
Theoretically, in addition to dynamics, statistical processes such as
particle emission and thermal fluctuations may be important.
In general, one needs
theoretical approaches accounting  for fluctuations in order to
predict the entire fission time distribution instead of the
average or most likely time.

\subsection{Mean-field theory}

The mean-field approximation provides the backbone of microscopic
nuclear theory for all but the lightest nuclei. In the context of
nuclear fission, the great advantage of the mean-field theory  is
that it is directly formulated in the intrinsic, body-fixed reference frame of the
nucleus, in which the concept of deformed nuclear shape and its
dynamical evolution  is naturally present.

Briefly, the self-consistent many-body wave functions are directly or
indirectly composed of Slater determinants of orbitals, with the
orbitals computed as eigenstates of one-body mean-field potential. If
the mean-field potential is determined by the expectation value of a
Hamiltonian in the Slater determinant, we arrive at Hartree-Fock (HF)
approximation. If a pairing field is included, we arrive at the
Hartree-Fock-Bogoliubov (HFB) approximation. As in electron density
functional theory (DFT) of condensed matter and atomic physics, the
Fock-space Hamiltonian is often replaced by an energy density
functional (EDF) defined through one-body densities or density matrices.
As is common practice in the nuclear physics literature, we will use
these notions interchangeably, where HFB and HF are used
to distinguish between nuclear DFT with (HFB) and without (HF)
treatment of pairing correlations. The use of an EDF instead of a
Hamilton operator sometimes necessitates to take different intermediate
steps in formal derivations, but leads to self-consistent equations that for
all practical purposes coincide with those of HF (or HFB if pairing is present).

Another approach in common use, the macroscopic-microscopic (MM) method, avoids the delicate issues
of constructing an EDF that reproduces the systematic properties of heavy
nuclei.  Here the basic properties of the nucleus are derived from
its size and shape, expressed in some parameterisation of the surface.
The orbitals are constructed with a potential
derived from the shape of the nuclear surface, and its energy is computed using the liquid
drop model together with shell corrections determined by the orbital
energies.  The first quantitative theoretical understanding of fission came from
this approach \cite{(Bra72a),(Bjo80a)}, see also its review in \citeasnoun{krappe2012},
and it has been successfully applied to calculate mass and
charge yields.

In HF and HFB, wave functions representing different nuclear shapes are
constructed by constraining the single-particle density matrix in some way.
This is often implemented by adding fields with Lagrange multipliers, but
it can also be done more directly; see {\Sec}~\ref{subsubsec:new}.
Typically, in nuclear DFT the nuclear shape is defined by several parameters that are
taken as collective variables.

\subsubsection{Potential energy surface}
\label{subsec:pes}

The potential energy surface (PES) represents the lowest possible energy
of the evolving system consistent with the specified values of
the collective variables.
As mentioned above,
the PES is generally multi-dimensional.
Although the PES alone does not suffice for predicting the dynamical evolution,
it is nevertheless very useful because its topography
makes it possible to understand and anticipate the main features of the
dynamics.
The local minima, saddle points, and the scission surface are key features
that often make it possible to predict isomeric properties, threshold energies, and fission fragment yields.

For a given point in the collective space,
the potential energy of the corresponding nuclear configuration
and its internal structure can be obtained either
by minimising the total energy in the CHF (Constrained HF) or CHFB (Constrained HFB) framework
or by calculating the MM energy for the specified shape.  The first
method results in an optimised shape within the given constraints
while the second method can miss aspects of the shape beyond the
defined shape parameterisation.
There are important consequences in both methods
for defining the collective space variables
and for the continuity of the resulting surface \cite{(Mol01a),(Dub12),SchunckDuke2014}.

While the standard PES describes the configuration having no excited orbitals
or quasi-particle excitations,
some approaches need energy in the presence of internal excitations.
In the MM method it requires the calculation
of shell and pairing corrections at finite excitation \cite{Ignatyuk1980},
while the self-consistent method may employ
a temperature-dependent DFT formalism \cite{EgidoRobledo2000,Pei2009,(She09),SchunckDuke2015,ZhuPei2016}.


\subsubsection{Other constraints in DFT}
\label{subsubsec:density-constrained}

The PES is usually presented as a function of a few multipole
moments in the CHF and CHFB framework, but multipole moments control
the shape only loosely and do not provide sufficient discrimination
between intrinsic configurations at large elongations.  When needed,
other types of constraints can provide
additional discrimination power.  For example one can define a neck-size
parameter to be added to the multipole moments \cite{Warda2002}.  More drastically,
the entire density distribution
$\rho(\mbox{\boldmath $r$})$ can be constrained.
Such a density-constrained method \cite{cusson1985,Uma85} has been used
successfully within TDHF (Time-dependent HF) approach to calculate heavy-ion interaction potentials
\cite{(Uma06),(Sim18)}.
For a sequence of shapes in the collision,
the instantaneous density of the evolving system obtained in TDHF is used as a constraint
for a static HF calculation,
yielding the lowest-energy configuration compatible with the constraint.
This eliminates both the collective kinetic energy and the internal excitation
and may therefore be interpreted as the potential energy.  While this
information is important for going beyond TDHF and TDHFB (Time-dependent HFB), there is no
simplification in the dynamics when taking the {\boldmath $r$}-dependent
density as a collective variable.  See Sect. \ref{subsec:dof} for
additional discussion of collective variables.

It is also possible to introduce constraints that depend more on the
wave function than on the shape.  In particular, one can get a high
discriminatory power in the space of axially symmetric configurations
by requiring a certain filling of the orbitals with respect to their
axial symmetry \cite{(Ber18)}. See also constraints pertaining to the
strengths of pairing correlations, discussed in {\Sec}~\ref{subsec:dof}.



\subsection{Time-dependent DFT}
\label{subsubsec:dynamics}

   The time-dependent version of HF is an established approach
to nuclear dynamics and has been extensively used to model
heavy ion collisions \cite{Simenel2012,(Sim18),sekizawa2019}.  In principle it can be easily generalised
to the HFB approximation, but one is only now reaching the computational
power to carry out calculations without introducing artificial
constraints and approximations  \cite{bulgac2016,(Has16),Scamps2017,Magierski2017,(Bul19a)}.  These approaches have an important
property that they respect energy conservation and the expectation
values of conserved one-body observables such as particle number.
Their strong point is that they usually give
a good description of the average behaviour of the system under
study.
Their weak point is that, since TDHF equations emerge as a classical
field theory for interacting single-particle fields
\cite{Kerman1976}, the TDDFT approach can neither  describe the
motion of the system in classically-forbidden part of the collective
space nor quantum fluctuations. As a consequence, the real-time TD
approach cannot be applied to SF theory. Moreover,  the fluctuations
in the final state observables, some being due to non-Newtonian
trajectories \cite{aritomo2014,Sadhukhan2017},  are often greatly
underestimated in time-dependent approaches.

\subsection{Beyond mean-field theory}

While the symmetry-broken product wave function of HFB already
provides a very good description for many properties, it is deficient
if a self-consistent mean-field symmetry is weakly broken. In such cases,
it is advisable to extend the method beyond a single-reference DFT. One way of
doing this is to use the small amplitude approximation to the TDHFB,
i.e., the  quasiparticle random phase approximation (QRPA). The QRPA
is a \emph{vertical} expansion that accounts for selected
correlations coming from excited states of the system. Another way of
enriching the DFT product state is through a multi-reference DFT
\cite{Bender2019book}. This represents a \emph{horizontal} expansion
\cite{Donau1989}. Two commonly used beyond-DFT methods
belong to this category. One is the generator coordinate method
(GCM). The GCM wave function is a superposition of single-reference
DFT states computed along a collective coordinate (or coordinates).
The second group contains various projection techniques, in which the
projection operation is applied to an HFB state in order to restore
internally-broken symmetries. The most advanced multi-reference DFT
approaches combine the virtues of the vertical and horizontal
expansion by employing the GCM based on the projected HFB states,
which often contain contributions from multi-quasiparticle
excitations.

\subsubsection{Generator coordinate method}\label{secGCM}

  A microscopic Hamiltonian treated in the CHF or CHFB approximations can
be mapped onto a collective Schr\"odinger equation (CSE) in the coordinates
defined by constraints.  This mapping is the essence of the
GCM.  Typically the mapping is carried out
using the Gaussian overlap approximation (GOA) to determine the kinetic
energy operator.  Examples of  such calculations for low-energy
fission  can be found in
\citeasnoun{(Gou04),Goutte2005,(Erl12e),Regnier2016,Zdeb2017,Tao2017,(Reg19a),(Zha19)}.
With several coordinates, the GCM produces much wider distribution in
the mass yields than can be realised in the evolution in time of a
single CHF or CHFB configuration.  On the other hand, the underlying
wave function is composed of zero-quasiparticle configurations and so
underestimates the non-collective internal energy.

To take into account non-adiabatic effects during the
fission process, the inclusion of excitations built on the
zero-quasiparticle vacuum becomes essential. Several
experimental observables attest to the importance of two-quasiparticle (2-qp)
excitations, which include the pair-breaking mechanism and the
coupling of pairs to the collective degrees of freedom.
From this point of view, the inclusion of explicit 2-qp components into the GCM wave function    is  of interest \cite{(Ber11a)}.
One of the
major advantages of the model is the nonlocal nature of the couplings
between collective modes and intrinsic excitations.
The development of this approach, however,
poses several problems related to the truncation of the 2-qp space;
 keeping track of  excitations along the collective path;   and evaluation of  overlap kernels.
So far, the model presented in \citeasnoun{(Ber11a)} has not yet been be  applied to fission problems.

\subsubsection{Projection techniques}
\label{sec:projections}

The nuclear Hamiltonian commutes with particle number,
angular momentum, and parity symmetry operations. The density functional of nuclear DFT is usually symmetry-covariant \cite{Carls2008,Rohozinski2010}.
Still, due to the spontaneous breaking of intrinsic symmetries in mean-field
theory, several symmetries are usually  broken in a nuclear DFT-modeling of fission.  There are
well-established projection methods to restore broken symmetries based on
the generalised Wick's theorem  \cite{(Man75a),StoitsovPNP,Bender2019book,(She19a)}
that have been applied to calculations of the fission barrier of $^{240}$Pu, either combining
parity and particle-number projection \cite{(Sam05)}, or combining angular-momentum and
particle-number projection with shape mixing \cite{(Ben04c)}.
The methods are straightforward in principle for models based on a Fock-space Hamiltonian.
Difficulties can arise in
EDF realisations of nuclear DFT, as discussed in \citeasnoun{Anguiano2001,(Dob07),(Ben09c),(Dug09),(She19a)}).
However, these problems do not concern the calculation of one-body observables such as the average particle
number in the fission fragments \cite{Reg19,(Bul19)}.

\subsection{Dissipative dynamics}\label{dissdyn}

While the self-consistent DFT  dynamics is very powerful, it largely
ignores the internal degrees of freedom that can bring large fluctuations
of observables and dissipate energy \cite{Kubo1966,Yamada2012}.  There are several ways that
the additional degrees of freedom can be taken into account in the
equation of motion.

A  simple diffusion master equation assumes the presence of first-order  time derivatives. This approach has been remarkably
successful in describing mass and charge yields \cite{(Ran11b)}.
While the utility of this ansatz has received
some support from recent microscopic calculations \cite{(Bul19a)},
its quantitative validity still needs to be derived.

More generally, one can consider time-dependent models that combine
time-even inertial dynamics with time-odd dissipative dynamics.  A
common classical formulation is with a multidimensional Langevin equation
\cite{sierk2017,usang2019}.  In this approach, the dissipated energy goes into
a heat reservoir characterised by a temperature. Recently, a hybrid Langevin-DFT approach has been applied to explain SF yields \cite{(Sad16a)}.
 While this is reasonable in a phenomenological theory, there is so far no microscopic
justification of this approach. It is to be noted, however, that  the predicted fission yield distributions  are found insensitive to large variations of dissipation tensor
\cite{(Ran11c),(Sad16a),sierk2017,matheson2019}.
The corresponding quantum dynamics requires an equation of motion for
the density matrix of the system.  One formulation is with the
Lindblad equation; see also \citeasnoun{bulgac2019fluc}.

\subsection{Quantum tunnelling}
\label{subsec:tunnelling}

Tunnelling motion in SF is usually treated via a quasiclassical, one-dimensional formula for
the action integral which is based on two main quantities that can be obtained in nuclear DFT: the PES
 and the collective inertia (or mass) tensor.
 The fission path is computed in a reduced
multidimensional space, using between two and five collective coordinates describing
the nuclear shape and pairing; see {\Sec}~\ref{subsec:dof}. The mass tensor requires
the assumption of a slow, near-adiabatic motion; see {\Sec}~\ref{subsec:coll_mass}. The pairing gap makes this
assumption most credible for even-even nuclei, but even in such systems one can
expect non-adiabatic effects due to level crossings \cite{SWilets1975,SWilets1975a,Strutinsky1977,Nazarewicz1993}.
The following questions are relevant for making
progress in SF studies.
\begin{description}
   \item[Generalised fission paths] Usually,  SF trajectories in the collective space are determined by considering several shape-constraining coordinates.
It is better to assume that the  collective motion happens  in a large space parameterised by  the Thouless matrix
           characterising a HFB state.
One approach to determine the collective path in that way has been proposed
in \citeasnoun{Marumori1980} and \citeasnoun{(Mat00)}. There the equations of motion have a canonical form (involving both coordinates and momenta), and constraining operators
are dynamically determined.
\item[Multi-dimensional WKB formula]
           The current barrier-penetration methodology is based on a
           minimisation of the collective action along one-dimensional paths,
           although our experience with above-barrier fission evolution
           suggest that the use of several degrees of freedom is important.
        It may be possible
           to generalise the one-dimensional quasiclassical WKB-like formula
          by a more general solution to a few-dimensional
           tunnelling problem  \cite{(Sca15)}.
   \item[Non-adiabatic effects] The admixtures of non-adiabatic
           states may be crucial to understand fission hindrance
           in odd nuclei.
The excitations to higher
           configurations can be induced by crossings of single-particle
           levels and by the Coriolis coupling; see {\Sec}~\ref{Landau-Zener}.
   \item[Instanton formalism]
           An alternative approach is provided by the formalism of
           imaginary-time TDHFB  \cite{Reinhardt1979,(Lev80a),(Pud87),Negele1989,(Ska08)}.
           Configuration mixing can be performed according to
           well defined equations, and spontaneous fission lifetimes
           could be determined  without having to
          define collective inertia.
           Non-self-consistent solutions using a phenomenological Woods-Saxon
           potential and omitting pairing have already been obtained \cite{(Bro18)}.
           If the simplified approach with pairing gives the proper order
           of magnitude for the fission hindrance and its weak dependence
           on particle numbers, the next step would be to incorporate the requirement
           of self-consistency.
\end{description}

\subsection{Level crossing dynamics}
\label{Landau-Zener}

In the original framework for a microscopic theory of fission
above the fission barrier,
Hill and Wheeler \cite{(Hil53)} proposed a model
based on time-dependent diabatic evolution of mean-field
configurations interrupted by possible jumps to other
configurations at the points of level crossings.  At those
intersections the probability to switch orbitals would be
computed by the Landau-Zener formula \cite{(Wit05)}.
This viewpoint has been pursued further in the later literature,
especially in the context of MM models \cite{SWilets1976,Norenberg1983,Matev1991} but the
challenges of implementing a microscopic theory has prevented
the actual calculation of macroscopic parameters such as friction
coefficients.

In the present era, computational resources are available
to carry out this program using DFT
and effective interactions to compute the interaction
matrix elements at level crossings.  Thus, we may now make
theoretical predictions of the balance between inertial
and dissipative dynamics that can be used as inputs to
more macroscopic models such as the ones solved with the
Langevin equation. The steps to carry out this program could
follow the strategy of the dissipative diabatic dynamics (DDD)
approach \cite{Norenberg1983,Norenberg1984,Berdichevsky1989,Matev1991,Mir14,Mirea2016}.
This would involve
the construction of diabatic PES, computing the interaction matrix elements between the
configurations that cross each other,  and
obtaining information about the time-dependence of the
motion along the path.  This can be achieved by
adding constraints on the velocity fields in the time-dependent evolution
of the configurations so that energy is conserved, see {\Sec}~\ref{subsubsec:new}.
With these additional tools one can explore the probability that there
will be some excitation of the nucleus along the fission path.
Namely, the probability of exciting the system from the adiabatic
path to a 4-qp excited state can be computed using the Landau-Zener
formula.
To get an actual dissipation rate, one would need to track a large
number of level crossing along the diabatic path.  There are
many issues that need to be studied carefully at this point such as (i)
non-orthogonality of the configuration basis; (ii) validation of the
level density against compound nucleus level density in the first
well; (iii) breaking down of the assumptions inherent in the Landau-Zener
formula at low velocities; and (iv) development of reliable statistical
approximations to deal with the large number of level crossings.

\subsection{Collective kinetic energy}
\label{subsubsec:ecoll}

The nuclear shape evolution generally rearranges the nucleons
and it is important to understand the associated collective kinetic energy.
Beyond the outer turning point,
while the electrostatic repulsion tends to accelerate toward the scission point, dissipative
couplings  damp the motion.
To connect with experiment, collective kinetic energy beyond the
saddle point is particularly important, because any relative motion at
scission adds to the fragment kinetic energy generated by the Coulomb repulsion
following scission.
While in an adiabatic description all the energy difference between
the saddle point (or the outer turning point for the low-energy fission)
and the scission point is converted into collective kinetic energy,
for strongly non-adiabatic motion, the system will irreversibly convert
most of that energy into intrinsic excitations,
endowing the nascent fragments with  little collective motion.

For the low-energy fission, where the motion is fairly adiabatic and the dynamics of the system is governed by a CSE,
the corresponding kinetic energy can be calculated
on the basis of the associated inertia tensor.
For a quantitative description of the collective kinetic energy
it is therefore essential to understand:
the relevant collective coordinates
(see {\Sec}~\ref{subsec:dof}); the  inertia tensor
(see {\Sec}~\ref{subsec:coll_mass});
the role of non-adiabatic effects
in general and during the descent to scission in particular
(see {\Sec}~\ref{subsec:dissipation}); and the role of dissipation, especially near scission.
In the time-dependent approaches,  the
kinetic energy can be  obtained by computing the collective current
as the local collective kinetic energy density $\propto \vec{j}^2$,
where $\vec{j}$ is the current density.

While most models agree that the pre-scission kinetic energy
forms only a small part of the final fragment kinetic energy,
there is no general consensus about its quantitative magnitude
\cite{(Bon07a),Bor08,SimenelUmar2014,bulgac2019fluc}. In general the TDDFT
calculations suggest that the evolution beyond the fission barrier is
strongly dissipative, and this impacts the predicted kinetic energy
\cite{bulgac2019fluc}.

It should be noted that TDHF models for high-energy fission are too diabatic, as the absence of pairing leads to
artificial fission hindrance \cite{(God15a)}.
The inclusion of pairing by allowing occupation number evolution
solves this hindrance problem \cite{Matev1991,Tan15,(Sca15b)}; see also
{\Sec}~\ref{subsec:lubricant}.

The calculation of collective kinetic energy and inertia for nuclei with an
odd number of protons and/or neutrons sometimes leads to diverging quantities.
While a solution to this problem is still missing, a natural strategy would be
to relax the adiabatic approximation. Note that this is also mandatory
when the two nascent fragments start accelerating close to the scission point. In this
context, the TDDFT is arguably the most suited method, since it naturally allows
the investigation of non-adiabatic effects in macroscopic transport
coefficients \cite{Tan15}.
The most important challenge for the TDDFT method is
the inclusion of dissipation along the fission path,
together with consistent fluctuations
in such a way that the fluctuation-dissipation theorem is satisfied.
The real challenge for this microscopic approach will be to properly describe
the energy exchange between collective and intrinsic degrees of freedom
 (see {\Sec}~\ref{subsec:dissipation} for more discussion).


\subsection{Approaches based on reaction theory}
\label{subsec:reaction}

Nuclear fission can be naturally formulated in the language of
reaction theory. Indeed, the SF process can be viewed as a decay of a
Gamow resonance, while the induced fission can be expressed as a
coupled-channel problem.
The description of
fission cross sections in
induced fission, for instance, is clearly in the domain of reaction theory.

There are two general frameworks for the
reaction theory of many-particle systems, namely $R$-matrix theory and
$K$-matrix theory.  The $R$-matrix framework has been extensively used
in the past to construct phenomenological treatments of
induced fission \cite{(Bjo80a)}.  But this approach is
not well adapted to microscopic calculations and has never
been applied at a microscopic level.  In contrast, the $K$-matrix theory
is closely allied with the configuration-interaction Hamiltonian approach
that has been  very successful in nuclear structure theory.
The $K$-matrix theory has been applied to a broad range of physics subfields,
but in nuclear physics only as a framework for
statistical reaction phenomenology \cite{kaw15}.
There are severe challenges to implementing the theory microscopically.
Some of these challenges are similar to those discussed in {\Sec}~\ref{Landau-Zener} in the context of microscopic DDD implementations.

First, one needs to construct a basis of non-orthogonal CHFB configurations that
effectively span the important intermediate states in the
fission dynamics.
This may be contrasted
with present  approaches that
rely heavily on an adiabatic approximation or TDDFT implementations.
Another  challenge is the need for microscopic
calculation of the decay width of internal configurations to continuum
final states of the daughter nuclei.  Tools based on the
GCM should be powerful enough to estimate the needed
widths \cite{(Ber19a),(Ber19)}.
It would take a large computational effort, and to date
no implementations of the GCM have be validated.
However, there  is some experience for
nuclear decays releasing an alpha particle \cite{(Bet12)} as well as
simple reactions involving light composite particles
\cite{(Wen17)}.

The $K$-matrix
reaction theory might be applied as a schematic model for testing
the approximations made in other approaches \cite{(Ber20)}.  In
particular, the importance of pairing in induced fission is not
well understood.  As mentioned in the next subsection, fission does not occur
on a reasonable time scale in pure TDHF at low energies;  adding pairing via TDHFB
lubricates the dynamics.

\subsection{Pairing as a fission lubricant}
\label{subsec:lubricant}

It is often said that pairing acts as a lubricant for fission. What
is meant by this assertion is that if pairing is removed from the treatment,
then the evolution from the ground state to scission
takes place through diabatic configurations which are often disconnected.
As a consequence, mean-field time evolution is sometimes unable to find the path to scission
\cite{(God15a)}. As realised early
\cite{(Mor74),Negele1989,Nazarewicz1993}, the pairing interaction
mixes those configurations and enables smooth transitions between
them \cite{NakatsukasaWalet}. The stronger the pairing is, the easier
these transitions are, and the faster fission occurs.

Pairing also plays an important role in the traditional WKB treatment
of SF.  The half life is proportional to the exponential
of the action,
 which in turn is proportional to the square root of the effective collective inertia.
The latter is proportional to the inverse of the square of the pairing gap,
so the stronger pairing correlations the smaller action and shorter half lives.
Indeed, numerous MM  studies
\cite{Urin1966,Lojewski1999,Staszczak1989} demonstrated that pairing
can significantly reduce the collective action; hence, affect
predicted spontaneous fission lifetimes. Implications of the pairing
strength being a collective degree of freedom for fission  are  very
significant, especially for the SF half-lives
\cite{Staszczak1989,(Giu14a),(Sad14a),(Zha16),Bernard2019}.


\subsection{Statistical excitation energy}
\label{subsec:temperature}

Apart from possible tunnelling, the fission path traverses the PES
at finite intrinsic excitation
energy.\footnote{For clarity, the excitation energy is the
difference between the total energy and the PES energy computed
in CHFB constrained to the same shape parameters.  The
collective kinetic energy is subtracted out to obtain the
intrinsic part.}  It can also be thought of as the energy of
the quasiparticle excitations in the fissioning nucleus.
Because the intrinsic energy is fairly high,
and the collective evolution is fairly slow,
the system has the character of a compound nucleus.
Therefore the intrinsic energy is often referred to as the \emph{statistical}
energy and characterised by a local temperature.
Any dynamical model of fission must therefore take into account statistical
excitation energy parameterised by a local temperature.
Furthermore, it is of interest to study how the fission process develops
as a function of total energy,
as is conveniently done in experiments inducing fission  by projectiles
at variable energies.
However, in the microscopic frameworks, the concept  of the finite temperature
is plagued by a number of conceptual and technical difficulties:

\begin{description}[align=left]

\item [Definition of Temperature] In the context of the
MM approaches, an effective, deformation-dependent
temperature can easily be defined following the recipes given in
\citeasnoun{Ignatyuk1980} and \citeasnoun{Diebel1981}. Given the local temperature at each point of
the collective space, one can construct an auxiliary potential energy
surface by damping the shell correction accordingly \cite{(Ran13)}.
This maintains a micro-canonical description of the process
where the total energy is constant, yet an effective PES exists and
can be used for dynamics. Such an approach is more difficult
in the DFT framework. First of all, many EDFs have an effective nucleon
mass well below unity, adversely affecting the relationship between
excitation energy and derived temperature.  Secondly, the connection between the
experimental excitation energy and the finite-temperature PES has not been clearly defined and,
in principle, calculations of dynamics should be carried out without its help
\cite{Pei2009,(She09),SchunckDuke2015,ZhuPei2016}.
In any case, it is important to have a good
definition of temperature
to describe
the disappearance of fission barriers, the increase in fluctuations, and
the damping of pairing and shell effects.

\item [Fluctuations] At finite excitation energy
the fissioning system displays statistical fluctuations
in addition to its inherent quantum fluctuations.
Therefore a large variety of outcomes is possible and, consequently, fluctuations of observables are significant.
As is obvious from the wide spreads in mass and charge yields, it is essential that the theoretical framework allows
the development of large fluctuations in the final outcome.
Moreover, because the possible final outcomes exhibit a very large diversity,
it is not feasible to express them as fluctuations
around an average.
Rather,
the only practical approach would provide an ensemble of outcomes
whose further fate (the primary fragment de-excitations process)
can then be followed individually and specific observables
can be extracted much as an ideal experiment would be analyzed.
This can be achieved in probabilistic treatments using
Monte-Carlo simulations.

\end{description}

\subsection{Coupling between degrees of freedom}
\label{subsec:dissipation}

The adiabatic approximation has often been employed to describe
spontaneous fission and low-energy induced fission.
In these formulations, the
coupling of the adiabatic
collective states to
the other internal degrees of freedom is a continuing challenge.
Nevertheless, it is important to assess how such couplings affect
decay rates and branching
ratios of the fission channel to other channels.
There are models available for the coupling, e.g.\ \citeasnoun{(Bri83),(Cal83)},
but they have never been validated in a microscopic reaction-theory
setting.  It is worth noticing, however,
that the classical Langevin equation can be derived using the
model by Caldeira and Leggett \cite{(Abe96a)}.
This fact might be utilised to extend the Langevin approach
to the quantal (tunnelling) regime.
That would be an important step for the theory of low-energy nuclear dynamics.

Another problem is that the adiabatic approximation breaks down at level crossings. In that situation,
a possible approach to treat dissipation is with the DDD approach (see {\Sec}~\ref{Landau-Zener}).

A challenge for microscopic theory is to include
adiabatic dynamics together with couplings to internal degrees of freedom.
Such a method should include a consistent treatment not only for
intermediate states but also for the collective inertia.
Current methods to compute inertial-mass tensor
rely on the adiabatic approximation
\cite{(Gia80b),(Mat00),(Hin07a),(Hin08),(Wen20)}.
A challenging problem is to develop a microscopic theory
for the large-amplitude collective motion
that takes into account non-adiabatic transitions.

Ideally, the dynamic equations would provide a time-dependent statistical
density matrix rather than the time-dependent wave function produced
by TDHF, TDHFB, etc.  An ambitious framework for such a theory has been
proposed in \citeasnoun{(Die10a)}.  It would require major additions to the present
coding algorithms as well as availability of high-performance computing
resource to implement.


\subsubsection{One- and two-body dissipation mechanisms.}
\label{subsubsec:diffusive}

In the theory of heavy ion reactions, it has been long recognised
that there are two distinct mechanisms that
arise in a semi-classical approach to dissipation \cite{(Sie80),(Ran84)}.  The one-body
dissipation operates at the level of TDDFT. It is fast
when it is present because the relevant time scale is
the time it takes a nucleon to transverse the nucleus.
The two-body dissipation is associated with nucleon-nucleon
collisions which are largely blocked at the Fermi surface;
its time scale is much longer.  Quantum mechanically, it
requires theoretical frameworks beyond mean-field theory, for
example, the inclusion of quasiparticle excitations in
the time-dependent wave functions.

In the semi-classical theory, the one-body dissipation can
be encapsulated in two formulas, the wall formula for
the internal dissipation in a large nucleus, and the
window formula for heavy ion reactions.  Both have been
used very successfully for many years.  However, the
assumptions required for the validity of the wall formula
may become questionable for low-energy fission dynamics:
time scales are long and shape changes are highly correlated
into low multipoles.

With the improvements in the computational capabilities
for carrying out TDDFT, it should be possible to map out
the region of validity of the semi-classical reductions
much better.  We now have credible evidence that the
one-body dissipation in a quantum framework is adequate
to dissipate the collective kinetic energy \cite{(Wad93)}, but still
not capable to produce a statistical equilibrium.

\paragraph{Fluctuations in collective variables.}

The presence of dissipation has two distinct but fundamentally related
effects on the evolution of the collective variables.
One is the average effect of the dissipative coupling
which acts as a friction force resisting the evolution;
this part is well described by the DFT.
The other arises from the remainder of the dissipative effect
which appears as a random force on the collective variables.
These two forces are related by the fluctuation-dissipation theorem \cite{Kubo1966},
often referred to as the Einstein relation.

As a consequence of the fluctuating force,
the system is continually faced with a multitude of trajectory branchings,
a situation that is very hard to encompass
within the usual microscopic frameworks.
That mean-field approaches, such as TDHF, are not suitable
for describing collective fluctuations has become especially apparent after
the advent of the variational approach by Balian and V\'en\'eroni \cite{Bal81}
who also proposed an alternative treatment of one-body fluctuations equivalent to
time-dependent random phase approximation (TDRPA) \cite{(Bal84a)}.
The practical applications of this  method to fission are still limited
\cite{(Sca15b),Wil18} and  further developments of the formalism are required.
A more radical approach would be to develop treatments that automatically
endow the collective variables with fluctuations by making their evolution
explicitly stochastic, as discussed in {\Sec}\ \ref{subsubsec:stochastic}.

\paragraph{Time-dependent generator coordinate method.}

A  quasiparticle HFB vacuum is
not expected to be a good approximation for  long-time evolutions.
A simple estimate leads to the conclusion that
the lifetime of such state is of the order of 100-200~fm$/c$,
whereas the time it takes from the saddle to scission
might exceed several thousand of fm$/c$.

For the long time evolution, the mean-field state is expected to couple to the
surrounding many-body states leading both to the breakdown of the
mean-field picture, and to a dispersion beyond mean field in the collective space
\cite{Goe80,Goe81}. This dispersion is usually
described by the TDGCM (Time-dependent GCM).

However, there are a number of limitations in current
applications of the TDGCM to fission that all employ the GOA.  For the moment, most
implementations assume that the collective motion stays in the
adiabatic PES. With this assumption, the manifestation of
non-adiabaticity, and henceforth a proper description of the transfer of
energy from collective motion to internal excitation, cannot be achieved.
Extending the TDGCM approach beyond the adiabatic limit \cite{(Ber11a),Reg19} to
incorporate dissipation and internal excitation, will require broadening the CSE
picture for the collective degrees of freedom; see, for instance,
\citeasnoun{(Die10a)}.

\paragraph{The internal equilibration process.}

Once the energy is transferred from the collective to the internal
degrees of freedom, it should be understood how the energy is subsequently
being redistributed so that internal statistical equilibrium is approached.

The onset of equilibration in interacting many-body systems
is a long-standing problem and several theories
have been proposed to treat this process \cite{(Abe96a),(Lac04),Sim10}.
In most treatments, it is assumed that repeated in-medium Pauli-suppressed
two-body collisions  lead the internal degrees of freedom
towards statistical equilibrium on a time scale that is relatively short
compared with that of the macroscopic evolution.

One example is the extended TDHF approach \cite{(Won78),(Won79),(Lac99)} or
its extensions based on the Bogoliubov-Born-Green-Kirkwood-Yvon hierarchy, generically called
time-dependent density matrix \cite{(Cas90a),Pet94}. These
approaches have rarely been used in nuclear reactions
\cite{(Toh02),Ass09} and specific technical problems seem to strongly
jeopardise the obtained results \cite{(Wen18)}.
However, a promising step forward has been achieved
with the purification technique, opening new perspectives \cite{Lac15b,Lac17}.

Apart from these technical issues,
this approach has the advantage that it leads
naturally to the Boltzmann-like description.
However, during fission,
especially as the system passes through the barrier region,
the excitation energy is sufficiently low to cause quantal and thermal
fluctuations to coexist.
This may lead to non-Markovian effects in the macroscopic evolution,
which obviously would complicate the treatment.
An extension of the TDGCM approach has been proposed for including
thermal fluctuations \cite{(Die10a)},
while quantum approaches treating the thermalisation process
have proven to be rather complicated. To circumvent them, approximate treatments
have been suggested,
including the relaxation-time approximation;
see \citeasnoun{Rei15}  for recent developments.


\subsubsection{Stochastic dynamics}
\label{subsubsec:stochastic}

Microscopic treatments of dissipation
are discussed in {\Sec}~\ref{Landau-Zener} and the previous section.
Such a level of detail can be avoided by a macroscopic transport approach,
treating its parameters
phenomenologically.  The equations here
describe the evolution of just a few collective properties,
typically the shape of the fissioning system,
as the initial compound nucleus evolves into two separate  fragments.
Because the retained collective degrees of freedom are coupled
dissipatively to the internal system,
the macroscopic evolution has a stochastic character
and the natural formal framework is the Langevin transport equation.
This treatment has been very successful \cite{sierk2017,usang2019}
in calculating a variety of fission observables.
A particular advantage of the Langevin dynamics is that it automatically
allows the collective trajectory to undergo dynamical branchings,
thereby making it possible for the system to evolve from a single  shape
to a large variety of final configurations.

Once the collective degrees of freedom have been identified,
the Langevin equation requires three ingredients:
the (multi-dimensional) PES,
the associated inertia tensor,
and the dissipation tensor describing the
coupling to the internal system and giving rise to both the collective
friction force and the diffusive behaviour of the collective evolution.
It is straightforward to apply microscopic theory to determine the first two of these key quantities.
For example, recent calculations of spontaneous-fission mass and charge yields
\cite{(Sad16a),Sadhukhan2017,matheson2019}, employed DFT
to obtain the PES and the inertia tensor as a function
of several collective coordinates,
then performed a WKB action minimisation for the tunnelling, and a subsequent
Langevin propagation until scission using a schematic dissipation tensor and random force.
Such a hybrid approach can be extended to the calculation of other fission
observables, such as the shapes and kinetic energies of the fragments.

The microscopic justification for the parameterised dissipation tensor
remains a problem.
As we have seen previously, TDDFT
includes one-body dissipation mechanisms. However, dissipation cannot
take place without fluctuations but it is not clear how to include
the fluctuations in the microscopic treatments.  Fluctuations inherent
in individual configurations of HF or HFB can be addressed by
the Stochastic Mean-Field
approach, which makes a  statistical assumption on the origin of
fluctuations, see  \citeasnoun{Ayik08,Lacroix14,tanimura2017} and references
therein. In this approach, the noise only stems from the initial
conditions. However, as it is well known in open quantum systems
theory, complex initial fluctuations can lead to a stochastic
dynamics with Markovian and non-Markovian noise continuously added in
time during the evolution. Understanding the connection between
initial fluctuation in collective space with the microscopic
Langevin approach on one side, and the link with current
phenomenological Langevin approaches on the other, should be addressed in the near
future.

In parallel, attempts have been made to reformulate quantum theories
leading to thermalisation as a stochastic process between
quasiparticle states \cite{Reinhard92,Lacroix06}
and important efforts are being made nowadays in condensed matter physics
to apply these methods \cite{Slama15,Lacombe16}. For the
moment, such reformulation have been essentially made assuming jumps
between Slater determinants, and equivalent formulations including
superfluidity is desirable for fission. A specific problem is again
that at very low excitation energy, stochastic approaches might face
the difficulty of exploring rare processes.


\subsubsection{Dissipation tensor}
\label{subsubsec:friction}

As is clear from the discussion in previous sections,
dissipation plays a key role in fission dynamics.
Langevin transport treatments of the collective evolution \cite{Kar01,sierk2017}
employ the simple wall and window formulas  (see {\Sec}~\ref{subsubsec:diffusive}) in various variants,
for example the chaos-weighted wall friction \cite{Pal98}.
In
many calculations, the dissipation tensor was phenomenologically adjusted
to reproduce experimental results.
In recent transport studies \cite{(Usa16),(Usa17)},
both the dissipation tensor and the inertial-mass tensor
were derived microscopically within the locally harmonic linear response
approach as outlined in \cite{(Iva99)},
but a validation of this method still remains to be carried out.

In general, one-body dissipation is rather insensitive to the
local nuclear temperature (whereas two-body dissipation is strongly energy
dependent, especially at low energy where the Pauli blocking is effective).
Recent studies \cite{(Sad16a)} have shown the importance of
dissipation in fission, even at energies relevant to spontaneous fission \cite{(Dag87)}.
As a consequence, the shape evolution acquires the character of
Brownian motion and many resulting observables,
most notably the fragment mass distribution, are rather independent of
the specific dissipation strength employed
\cite{(Ran11c),sierk2017,(Sad16a)}.

One observable that is somewhat sensitive to the
dissipation strength is the final fragment kinetic energy,
a quantity that has proven to be difficult to treat reliably in models.
By contrast, the time elapsed from the crossing of the fission barrier
until scission is quite sensitive to the dissipation,
being roughly inversely proportional to its strength.
However, this quantity is difficult to measure directly,
though somewhat equivalent experimental information can be obtained
from quasi-fission processes \cite{Wil18,(Ban19)}.

It is an important challenge to derive the dissipation tensor
from microscopic models.  For this,
the TDDFT method (including pairing) might be a suitable tool.
In its basic form, by energy conservation and by the knowledge of the
kinetic energy and excitation of the nascent fragments after scission,
one can determine the total energy dissipated from the initial condition.
(The excitation of the nascent fragments is initially partly given in the form
of distortion energy which will gradually be converted to additional
internal excitation as the fragment shapes relax to their equilibrium forms.)
Two existing approaches might be useful for obtaining
information on dissipation in TDDFT. The first is
the density-constrained TDHF method of {\Sec}~\ref{subsubsec:density-constrained}.
More systematic application of this approach to disentangle the
collective energy from the excitation energy without imposing the
adiabatic approximation is desirable. An alternative approach,
called  Dissipative-Dynamics TDHF, consists in making a
macroscopic mapping of the collective evolution \cite{Was08,Was09}
which, however, requires a somewhat ambiguous choice of
the relevant collective coordinates.  This approach has not yet been
applied to the fission problem,
although a first step in this direction has been made \cite{Tan15}.


\section{Many-body inputs}
\label{sec:inputs}

The treatment of collective nuclear dynamics in fission
requires a variety of inputs that can be obtained from many-body theory.


\subsection{Collective degrees of freedom}
\label{subsec:dof}

The starting point in the study of large amplitude collective dynamics
is the identification of the degrees of freedom to go into the
equations of motion.
Although collective coordinates are not direct observables,
they are treated as physical degrees of freedom; they are
required for the construction of the PES
as well as the associated inertia- and dissipation tensors.
To achieve a satisfactory description of fission observables,
such as the fragment mass distribution or the total fragment kinetic energy,
it is essential to include a sufficiently rich set of collective coordinates.
For example, even though the principal fission degree of freedom
is the overall elongation, it is necessary to also include a shape
coordinate
breaking reflection symmetry to obtain realistic fragment mass yields.
It was argued long ago \cite{NixNPA130} that a reasonable description
must use a minimum of five degrees of freedom,
namely overall elongation,
necking,
reflection asymmetry,
and the shapes of the two emerging fragments.
It appears that an overall intensity of pairing correlations, treated
as a degree of freedom, should also be added to this list
\cite{Staszczak1989,(Giu14a),(Sad14a),(Zha16),Bernard2019}.
However, the number used in actual studies is often smaller,
primarily due to computational considerations.

Within the framework of MM treatments,
the principal collective degrees of freedom
are those characterising the nuclear shape.
A variety of shape families have been employed.
Probably the most widely used are
the three-quadratic-surface parameterisation in \citeasnoun{NixNPA130},
and the parameterisation of  \citeasnoun{(Bra72a)}, which have three parameters.
A detailed discussion of the advantage of one particular
parameterisation over another can be found in \citeasnoun{(Mol09a)}.
However, even five shape degrees of freedom may not always be sufficient.
For example, triaxial shape deformations are often important in the region of the first saddle.

Self-consistent treatments based on nuclear energy density functionals
have used multipole moments of the matter distribution as constraints
to play the role of collective coordinates \cite{krappe2012,younes2019}.
The primary collective coordinates employed in such studies
are the quadrupole moments $Q_{20}$ and $Q_{22}$
used to control the overall distortion and triaxiality of the system, respectively,
the octupole moment $Q_{30}$, used to control its reflection asymmetry, and the neck parameter or  the hexadecapole moment $Q_{40}$. An interesting possibility is to generalise the use of
a set of multipole moments as the constraining operators
by  using the density distribution itself, see {\Sec}~\ref{subsubsec:density-constrained} for further discussion.

It is important to recognise the principal difference between
the use of the nuclear shape as a (multi-dimensional) collective variable,
as is done in the macroscopic-microscopic approaches, and the
use of a set of density moments, as is being done in the microscopic treatments.
Whereas the former approach calculates the properties of the system
having the specified shape,
the latter automatically performs energy minimisation
so the system being treated is the one having the lowest energy
subject to the specified moment constraints.
Consequently its shape (or more generally: its matter distribution)
is not under complete control.
As discussed in {\Sec}~\ref{subsec:pes},
the  self-consistent  density distribution may exhibit discontinuities
as the moments are varied smoothly as a small change in the constraints might
cause the new minimal state to have a quite different spatial appearance.
This problem is particularly severe near the scission point, where there
might be a major reorganisation of orbital fillings.
A recent detailed study of this problem \cite{(Dub12)}
developed diagnostic tools for identifying its presence
and demonstrated how additional constraints could help.
In any case, no set of collective coordinates were found that
could eliminate the problem entirely.
It is therefore clear that at least three collective constraints are needed
to mitigate such discontinuities.

Fluctuations of the pairing field have also been used as collective coordinates \cite{Staszczak1989}, see {\Sec}~\ref{subsec:lubricant}.
Here, a constraint on the dispersion in particle number  $\langle N^2\rangle
- \langle N\rangle^2$ is imposed to control the strength of the pairing field
\cite{(Vaq11),(Vaq13)}.
Studies of fission dynamics
have shown that the coupling between shape and pairing degrees of freedom
has in fact a significant effect on the collective inertia and, therefore,
on the dynamical paths in the collective space.  In particular,
it may have a pronounced influence on  spontaneous fission half-lives
\cite{(Sad14a),(Zha16),Bernard2019}.
Pairing coordinates  may also be  important for the odd-even staggering
in the fission yields \cite{Mir14,(Rod17)}.
Induced fission is traditionally treated in a finite-temperature framework,
where pairing is quickly quenched by the statistical fluctuations.
Here, again, the dynamical treatment of pairing
could substantially change the picture.

Generally, the introduction of additional collective coordinates
increases the required numerical effort significantly.
Nevertheless, for more refined descriptions,
there is a need for a few additional collective variables
that are not shape-related.
One is the projection of total angular momentum on the fission axis,
usually denoted by $K$ \cite{(Nad12a),(Ber18)},
which affects the angular distribution of the fission fragments.
In a recent study, the configuration space was constructed
in the HF approximation using the $K$-partition numbers
as additional constraints \cite{(Ber18)}.

Another additional collective degree of freedom is related to the isospin.
Except for TDHF, TDHFB, and DFT-Langevin, fission treatments have usually assumed that the fragments
retain the same proton-to-neutron ratio as that of the mother nucleus.
While some progress has recently been made in incorporating
this degree of freedom into the MM treatments
\cite{(Mol14),(Mol15a),MollerEPJA53},
further developments are still needed.

A near-term challenge will be to take advantage of newly available
extensive computing resources and expand the space of collective
coordinates,
with the aim of obtaining a more realistic description of the evolution
of the fissioning nucleus into fragments,
especially in the region where nascent fragments appear near and beyond scission.


\subsection{Collective inertia}
\label{subsec:coll_mass}

The ATDHFB (Adiabatic time-dependent HFB) and GCM+GOA formalisms are often applied to derive collective
inertias for
the CSE.
In the ATDHFB this requires the inversion of the full linear response matrix.  From a computational point of
view, this is a daunting task that has been often alleviated by imposing various
approximations \cite{(Sch16b)}.
Typically, fission
calculations rely on the ATDHFB inertias within the so-called non-perturbative
cranking approximation, where the non-diagonal terms of the linear response
matrix are neglected and the derivatives of the generalised density matrix with
respect to the collective variables are computed numerically \cite{(Bar11)}. Very
recently, both the exact and non-perturbative cranking GCM+GOA inertias have
been computed for the first time \cite{Giu18}, showing that the non-perturbative
cranking ATDHFB inertias can be reproduced even without the inclusion of
collective momentum variables.

In the TDGCM framework, the expression for the collective kinetic energy can
be obtained using either the GCM or the ATDHFB formalism.  While the latter
approach leads to the physical inertia in the case of translational
motion \cite{Rin80}, the GCM approach may be incorrect if the
conjugate collective variables are not included as collective degrees of freedom.
This requires doubling the dimensionality
of the collective space and in practice this is rarely if ever done \cite{Goe80}.
To obtain a more realistic inertia, the ATDHFB expression is sometimes
used in the GCM approaches. The possibility of using fully consistent GCM
with pairs of collective variables would be desirable in the future.

Given the current status, several aspects should be addressed in order to reduce
the source of uncertainties in the estimation of collective inertias.  Among the
most impelling ones is the calculation of the exact ATDHFB inertias.
This is desirable because,  according to the instanton formulation
(see \citeasnoun{(Ska08)} and {\Sec}~\ref{subsec:tunnelling}), it is
the ATDHFB that provides a compatible framework to tackle the problem of
nuclear dynamics under the barrier. The
full linear response matrix has been inverted in \citeasnoun{(Lec15)} under some approximations
but  this method has not yet been extended to
fission studies. Alternatively, one could try an approach along the lines of the Finite
Amplitude Method (FAM) \cite{Hin15}, where rather than inverting the linear
response matrix itself one computes its action on the time derivative of the
density matrix entering in the expression of the collective inertias.
Such an approach has already been proposed and tried long time ago
\cite{(Dob81a)}, and, as advocated in \citeasnoun{(Dob19a)},
the time is ripe to start implementing it routinely in all ATDHFB calculations.
Regardless of the practical implementation, the estimation of the exact ATDHFB
inertias  is a crucial step to
understand the validity of the non-perturbative cranking approximations, which
will reduce the uncertainties related to the collective inertias and bring a
sounder estimation of collective kinetic energies and in the general adiabatic
description of the fission process.

When it comes to non-adiabatic formulations, collective inertia can
be derived within the DDD formalism \cite{Mir19}.  For low collective
velocities, the DDD inertia reduce to the  cranking expressions.


\subsection{Collective dissipation}
\label{subsec:coll_diss}

In most treatments of the fission dynamics based on microscopic theory,
it has been assumed that the collective degrees of freedom are well
decoupled from the intrinsic degrees of freedom,
usually referred to as the adiabatic assumption.
Unfortunately,  the nuclear $A$-body wave function of the nucleus cannot, in general,
be expressed in terms of slow and fast components. Indeed, the
typical time scale of nuclear collective modes is only slightly
greater than the
single-particle time scale \cite{Nazarewicz2001}. In the context of fission, the
adiabatic approximation is questionable
as the collective motion is highly dissipative
\cite{Blo78,bulgac2019fluc}, see {\Sec}~\ref{dissdyn}.

There is therefore an urgent need for addressing the collective dissipation
within a microscopic framework.
While this presents a significant computational undertaking,
the most immediate task consists in deriving the appropriate expressions
for the dissipation in the particular microscopic model employed,
a problem that is still quite unsettled \cite{(Bar78c)}. Another challenge is to
identify high-quality  fission data that will constraint the dissipation
tensor.


\subsection{Level densities}
\label{subsec:levels}

The
nuclear level density is a key ingredient of the Hauser-Feschbach
statistical theory of nuclear reactions. Modeling many aspects of
fission reactions rely on this type of statistical reaction theory:
a first example is nucleon-induced fission, where the capture of the
projectile by the target and the fission of the resulting compound
nucleus are treated as a two-step process. Another example is the prompt
de-excitation of the nascent fission fragments, which  can be treated
as compound nuclei undergoing statistical decays. Especially for
applications in nuclear astrophysics, such as the calculation of
fission transmission coefficients and fission yields, reliable
predictions of nuclear level densities over a broad range of excitations
for a large region of nuclei are desired.

Three main classes of nuclear level density models exist: analytical
models (such as the back-shifted Fermi gas),
configuration-interaction methods, and combinatorial model. Due to
their simplicity, analytical models are often used in reaction codes,
but they do not account for specific nuclear structure effects to a
satisfactory degree. While in principle very powerful,
configuration-interaction shell-model methods have so far been
applied only to a limited number of light and medium-mass nuclei
because of their computational complexity. In contrast, the
shell-model Monte Carlo (SMMC) approach is capable of calculating
level densities of heavy nuclei and was applied to nuclei as heavy as
the lanthanides \cite{(Alh08)}.  SMMC level densities have the
advantage that they include contributions from both intrinsic and
collective excitations. However, application of SMMC across the
nuclear chart will require large computational resources.

Combinatorial models do not suffer from this hurdle and have been
applied on the scale of the nuclear chart. These are usually based on
the microscopic single-particle levels (provided by DFT calculations,
microscopic-macroscopic approaches, or analytical optical potentials)
from which the many-quasiparticle excited states are obtained
after pairing has been included.
The level density obtained by such combinatorial counting
must be augmented by the effect of excited states
that are mostly collective in nature.
Most important is the appearance of rotational bands for deformed nuclei
which may increase the level density by more than an order of magnitude
even at moderate excitation energies.
Even though this effect is very important, most treatments have long
included it only by means of an empirical formula
based on the moment of inertia of the nucleus \cite{(Bj73)}.
However, more recent approaches have considered each individual
many-quasiparticle excited state to be a rotational bandhead
\cite{UhrenholtNPA913},
thus avoiding the introduction of adjustable parameters.
Collective vibrations have also been included \cite{(Hil12),UhrenholtNPA913},
but these are most often neglected as they have been found
to have only a small impact at low excitations
compared to the rotational enhancement.

An additional aspect of the modelling of these collective enhancements
is their dependence on the nuclear shape.
For example, photofission rates are sensitive to the ratio
of level densities at the ground state and at the fission saddle point.
Many transport models use the level density
(as a function of the collective coordinates)
to relate the local excitation energy to a local temperature.  Furthermore,
recent transport treatments of fission have employed shape-dependent
level densities to guide the nuclear shape evolution \cite{(War17)},
an approach that automatically takes account of the gradual decrease
of pairing and shell effects at increasing excitation.
The effect of this energy dependence is often emulated by using
a phenomenological damping function for the level density.
Finally, shape-dependent microscopic level densities are also important
for the division of the internal excitation energy
between the pre-fragments at scission \cite{(Alb20)}.

Statistical quantities are important in many aspects of fission,
and microscopic theory is needed to go beyond the current empirical
modelling of their dependence on shape and other variables.
Particularly challenging is the problem of calculating
shape-dependent level densities
with a proper description of the gradual erosion
of the shell effects with increasing energy.


\section{Initial conditions}
\label{sec:entrance}


\begin{figure*}[!htb]
\begin{flushright}{\includegraphics[width=0.85\linewidth]{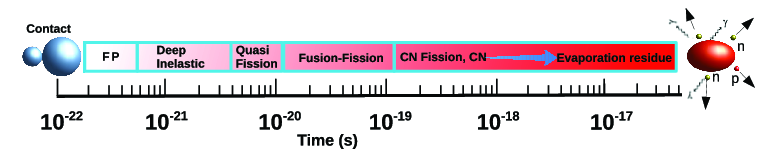}}\end{flushright}
\caption{\label{time-scales}Time scales for various reactions that may
induce fission in large nuclei. FP means fast probes and CN stands for compound nucleus.}
\end{figure*}

Nuclear fission can proceed from a variety of initial states; see  {\Sec}~\ref{subsec:sf}.
In addition to spontaneous fission,
the process can be induced by a variety of nuclear reactions
that lead to the initial state of the fissioning (mother) nucleus.
Figure~\ref{time-scales} displays some
characteristic time intervals for the preparation of the initial state.
The reaction types are roughly ordered
by the amount of energy they may bring into the compound system, with the
most energetic reactions (``Fast probes'') on the left.  On the right,
labeled ``CN fission'', are the more gentle probes such as neutron capture
that proceed through the compound nucleus.

\subsection{Neutron-induced fission}
\label{subsec:neutron-induced}

Due to its importance in applications,
low-energy  neutron-induced fission is probably the best
experimentally studied and phenomenologically parameterised fission
process. Proceeding through narrow neutron resonances, it
creates the nucleus in a long-lived excited state that has enough
time to thermalise the absorbed energy, thus forming a compound nucleus.

The incident neutron creates a compound nucleus at an excitation
energy that can range from somewhat below the barrier to energies above the barrier.
As the neutron energy is raised, the resulting excitation may
make it possible for the nucleus to evaporate one or more neutrons
before fissioning (referred to as a multi-chance fission).
At still higher energies, non-equilibrium emission grows important
and a thermalised compound nucleus is established only after the
loss of one (or possibly more) nucleons.

The representation of the initial compound state
in terms of elementary excitations is impractical
because the level density is prohibitively high.
Thus a direct description in terms of QRPA modes, GCM excited states, etc.,
would not be feasible and it may be preferable to adopt an approach based on
statistical quantum mechanics.
However, for the neutron-induced fission close to the neutron drip line, where level
densities at the neutron separation energies are small, the process
could still be dominated by the direct component rather than the compound one.

 Besides the energy, all the quantum numbers of the
formed compound nucleus may affect the subsequent fission process,
especially when the energy is close to  the barrier height.  This level of
detail is retained in the $R$-matrix theory, discussed in {\Sec}~\ref{subsec:reaction}.


\subsection{Fission induced by fast probes}
\label{subsec:fast-probes}

Fission can be induced by fast probes such as photons (photofission),
charged particles, and high-energy neutrons. Surrogate reactions,
such as fission following multi-nucleon transfer reactions, are also
included here.
In these processes, the nucleus is created in an excited state
above or below the fission barrier. This state
may exhibit specific well-defined structures,
such as the giant dipole resonance, which have substantial widths.

Present theoretical descriptions of fission induced by fast probes
most often assume the creation of a compound nucleus
at a given thermal excitation energy
(cf~the process of neutron-induced fission discussed in
{\Sec}~\ref{subsec:neutron-induced}).
However, for fast probes such an assumption might be ill-founded
because the nuclear system may not have sufficient time
to thermalise before undergoing fission.
Therefore, a non-thermal description of fission at high excitation
energies is very much desired \cite{(Dob19a)}.

This becomes increasingly important at higher energies
where pre-equilibrium processes play an increasingly significant role
and may lead to the emission of one or more nucleons
before equilibrium is reached.
Moreover, as the excitation energy of the compound nucleus is increased,
neutron evaporation competes ever more favourably with fission and,
as a result, multi-chance fission is likely to happen.

\subsection{Formation by electromagnetic and weak-interaction fields}
\label{sect:EW}

\subsubsection{Photofission}

In photofission, a nucleus decays through the fission channel
after absorbing a high-energy photon -- a $\gamma$-ray. The characteristics of
the excited state resulting from photo-absorption -- the initial state for the
fission process -- determines the evolution of the system, for instance, by determining
whether enough excitation energy is available to surmount the fission barrier.
Thus, the knowledge of excited states above both the ground state (for fissile
nuclei) and shape isomers, as well as multipole transition
probabilities between these states, is in principle needed to model
photo-absorption as a function of the photon energy. If the
photon is absorbed through the dipole operator on an even-even nucleus,
the angular distribution of the fission fragments gives information about
the  mixing of the $K$ quantum number in the fission process.  In general
and outside of the giant dipole absorption peak, theories such as the QRPA
are needed to sort out  the multipoles.

\subsubsection{Coulomb excitation}
 Another electromagnetic excitation method to study
fission of heavy nuclei in a relativistic accelerator beam is
Coulomb excitation.  Here, the process can be
treated as excitation by virtual E1 photons, so the considerations in
the previous paragraph apply.  While the energy transferred  is not precisely known, the theory for its distribution is well established.

\subsubsection{$\beta$ decay and electron capture}
Fission of nuclei far from stability can sometimes
be studied when the nuclide is formed by $\beta$ decay of a progenitor
nuclide.
In terms of the excitation energy, $\beta$-delayed fission is
intermediate between SF and Coulomb-excitation induced fission.
Importantly, this process makes it possible to study low-energy
fission in proton-rich heavy nuclei that are  not  accessible by
other techniques \cite{Andreyev2013}.
As in Coulomb excitation, the excitation energy given to the
nucleus is not known precisely.  Thus the theory of $\beta$-decay strength
function is required to model the whole process.  In this regard, the
QRPA (in its charge-exchange formulation) is very valuable.

The process of $\beta$-delayed fission also plays an important role in nucleosynthesis,
because it helps to terminate the rapid-neutron-capture process.
Fission may occur from the compound nuclei created by neutron capture
or from the $\beta$-decay daughters of those nuclei \cite{(Mum18)}.
The latter can happen whenever the $\beta$ decay populates a daughter state
with an excitation energy above (or near) the height of the fission barrier.
Since it is important to know the spin and the parity of the initial state
before fission, the description of $\beta$-delayed
fission requires a microscopic model of the charge-exchange
process
to provide $\beta$-strength distributions; for the  recent QRPA work see
\citeasnoun{(Mus14),(Mus16),(Sha16)}.
The QRPA applications used to describe $\beta$-decay are often limited to
allowed transitions. Thus it would be necessary to extend many
current QRPA codes to enable computation of all possible final states in daughter nuclei.

Because  $\beta$-delayed fission often involves odd-odd
nuclei, one should employ a formalism that can be extended to such
systems without introducing any additional approximation. Therefore, both the underlying HFB solver
as well as the QRPA implementation should break time-reversal symmetry, that is, extend
beyond the equal filling approximation. This last point is essential to
differentiate between low- and high-spin states in odd-odd nuclei, and thus
distinguish between decays from potential isomeric states and the ground state.
Once the fissioning  daughter state
has been determined, one should be able to calculate the corresponding
potential energy surface for the particular energy, spin, and parity.


\subsection{Heavy-ion reactions}
\label{subsec:fus-fis}

In the search for superheavy nuclei, the experiment uses a
heavy-ion reaction to fuse together two large nuclei, hoping that the
combined system equilibrates and then decays as a compound nucleus.
Cross sections can be estimated for this reaction mechanism, but a
crucial ingredient is the probability to form a compound nucleus.  The
reaction is called fusion-fission in that case; if there is no equilibration
it is called quasifission.
The understanding of this distinction requires a combination
of statistical and truly dynamical approaches which are not necessarily confined to a
collective subspace.
Quasifission leads to the formation of products that may have
similar properties to fission products, but are produced without the formation
of compound nucleus.
Fusion-fission occurs after the formation of a composite system which
fissions due to its excitation, resulting in a fragment distribution
that is peaked at equal mass breakup of the composite system. This
difference in fragment distributions indicates that quasifission is
the faster process and corresponds to a system that is not yet
fully equilibrated. As a dynamical process, quasifission is amenable
to a description using the TDHF
approach \cite{(Sim18)}. A number of TDHF studies of heavy-ion reactions have been reported in recent
years \cite{wakhle2014,(Obe14),(Uma16),(God19),sekizawa2019,(God20)}.
In general, the TDHF results agree well with the experimental quasifission yields,
and shed light on some of the underlying reaction
dynamics in relation to target/projectile combinations.

Quasi-fission and fusion-fission
could be used to map out the non-adiabatic collective landscape between the fusion entrance channel
and the fission exit channel.
The calculated time scales indicate that while fast
quasifission events dominate, much slower events resulting in a fracture with equal mass fragments have also been observed.

One of the open experimental questions is how to distinguish quasifission from
fusion-fission. This is important for calculation of the evaporation residue formation
probability in superheavy element  searches.
A collaborative effort between theory and experiment is needed to find ways to address
these issues. One may try to ``calibrate'' the experimental quasifission yields with the help of
theoretical simulations thus allowing the extraction of the fusion-fission yield. Study of
angular distributions (now routinely measured with large angular acceptance detectors \cite{(Ban19)})
may be one of the ways to approach this task.

Theoretical studies of quasifission have taught us that the dynamics may be dominated
by shell effects \cite{(Sim18),sekizawa2019}.
Despite the apparent strong differences between fission and quasifission, it is interesting to note that
similar shell effects are found in both phenomena \cite{scamps2018,ScampsSimenel2019,(God19)}. Quasifission
can then potentially be used as an alternative mechanisms to probe fission mode properties. For instance,
this could provide a much cheaper way than fusion-fission to test the influence of the $^{208}$Pb shell effects in
super-asymmetric SHE fission. Note that this approach would only provide information on the properties of fission
modes (mass asymmetry, TKE, excitation energy), but not directly on their competition. Indeed, the latter is likely to be
determined near the saddle point, a region of the PES which is not necessarily explored by the quasifission paths.


\section{Forces for fission dynamics}
\label{sec:EDF}

The collective fission dynamics can be understood as a balancing of
three different types of forces: the driving forces arising from the
generally multi-dimensional potential energy of deformation
of the fissioning system,
the inertial forces caused by the macroscopic rearrangement
of the nucleons associated with the change of the collective coordinates,
and the dissipative forces arising from the coupling of the considered
collective coordinates to the remainder of the nuclear system.
In a microscopic approach  these fission-driving forces are derived
from the effective inter-nucleon interactions, which are optimised to
selected data.


\subsection{Energy density functional}
\label{subsec:params}

In a microscopic approach to fission, the effective inter-nucleon interaction or energy density functional
is the only ingredient of the theory that includes adjustable
parameters. Therefore, the choice of a functional ultimately
determines the quality of the microscopic description of phenomena
related to fission, and the level of quantitative agreement with data.

Several types of EDF have been proposed over the years,
many of which have also been applied to fission. These functionals
can be non-relativistic or relativistic (or covariant), and this choice leads to different equations of
motion for nucleons; they can be functionals of local or non-local densities; they can be
strictly defined as the expectation value of a corresponding generating many-body operator or not;
and finally the couplings (parameters) can be constants or include a medium (density) dependence.

The two most widely used non-relativistic EDFs \cite{(Ben03e),(Sch19b)}
are the finite-range Gogny EDF, which is constructed including the HFB
expectation value of a density-dependent interaction,
and the  Skyrme EDF, which includes momentum- and density-dependent
zero-range terms in the interaction.
Other types of
local non-relativistic EDFs that were recently developed
and applied to detailed
studies of fission processes are the  Barcelona-Catania-Paris-Madrid
\cite{(Bal13)}, SEI \cite{(Beh16)},
and SeaLL \cite{(Bul18)} EDFs.
Similarly, there are several varieties of relativistic  EDFs in use
\cite{(Sch19b),(Agb17),Agb19}, either with finite-range (meson-exchange) or contact interaction potentials, with
non-linearities in the meson and/or nucleon fields, or including density-dependent couplings.
Two relativistic point-coupling (contact) functionals, in particular, have successfully been applied to studies
of fission dynamics: PC-PK1 \cite{(Zha10)} and DD-PC1 \cite{(Nik08)}.

One important challenge is to increase the predictive power of novel nuclear
EDFs compared to traditional functionals such as Gogny or Skyrme, which apparently cannot be
improved further \cite{(Kor14)}. For instance, the density-matrix expansion \cite{(Car10c),(Geb10),(Geb11),(Sto10)} can be
used to construct nuclear EDFs that are guided by first principles \cite{(Dyh17),(Nav18),(Zha18a)}.
Extensions of time-tested EDFs have been subject to recent studies.  For
example, higher-order
gradient terms have been added to the Skyrme EDF \cite{Carls2008,(Dav13a),(Bec17a)}.
The Gogny family of functionals have been extended to include additional
density-dependence and tensor interactions
\cite{(Cha15),(Pil17),(Ber20a)}. One hopes that such extensions would augment the
parameter space to be optimised for a better
description of fission properties.

Over the past decade it has been realised that exact projection techniques (see
 {\Sec}~\ref{sec:projections}) and
exact GCM
are ill-defined for EDFs that are not strictly constructed
from an effective Fock-space Hamiltonian
\cite{Anguiano2001,(Dob07),(Lac09),(Ben09c),(Dug09),(Rob10b)}.
The basic dilemma that one faces in this context is that a suitable form
of an effective Hamiltonian
that reaches the
descriptive power of conventional EDFs has not yet been
identified. As a first step in this exploration, a
scheme for a systematic construction of flexible two-body
interactions by combining finite-range Gaussians and gradients, has been
proposed \cite{(Dob12),(Ben17a)}. Limiting oneself
to a two-body interaction, however, will inevitably lead to an
unrealistically small effective mass \cite{(Dav18)}, such that one
always has to add three-body, and perhaps even higher, interactions.
The computationally simplest form of such terms is provided by
contact three-body forces with gradients \cite{(Sad13b)}. It
turns out, however, that when added to two-body interactions of
various forms, they do not offer sufficient flexibility, which makes
this quest even more challenging.

Many advanced methods
for beyond-DFT modelling of fission dynamics often
include explicit correlation energies that were implicit in the effective
interaction obtained from the EDF optimisation.  This inconsistency may
degrade the
descriptive and predictive power of the model and should be avoided. A
better strategy would be to optimise the EDF parameters using  data sensitive to large deformations.
This issue is most obvious in the case of corrections for quantal zero-point
motion related to symmetry breaking and shape fluctuations, such as those
for the centre-of-mass, rotational, and shape-vibrational motion. For instance,
the inertia that determines the
former is the mass number $A$, which becomes ambiguous whenever
one considers the separation of a single nucleus into fragments
\cite{(Goe83),(Ska06)}. The rotational correction increases
with deformation and therefore lowers fission barriers, etc.
To further complicate matters, one form of quantal
correction is transformed into other forms when changing deformation
\cite{(Goe83),(Ska06)}, such that from this point of view many
quantal corrections have to be treated simultaneously.
The same considerations also apply to exact projections and full GCM.

Static and dynamic pairing correlations play a crucial role for
the calculation of deformation energy surfaces, the dynamic fission path, and collective inertia.
This means that the pairing part of the effective interaction or EDF might have to be
tailored in such a way to reproduce both ground-state properties and selected features that determine fission data, see {\Sec}~\ref{subsec:optim}.

\subsection{Optimisation strategies}
\label{subsec:optim}

Once the form and the framework for which the parameters of an EDF
are to be adjusted are decided, the next question concerns the
selection of fit observables. Most of the fission
observables (lifetimes,
fission fragment distributions, \ldots) are computationally expensive and cannot be
systematically considered during the optimisation. Therefore, one has
to identify properties that encapsulate
the essence of the relevant physics probed by fission and, at the same time,
can be computed in a reasonable time.

First of all, the EDF has to be capable of describing the structure
of the initial state of the fissioning nucleus
and the final state of the fragments. At low excitation energy, the
requirements for this are the same as for standard nuclear structure
applications.
One of the most important constraints on the EDF specifically relevant for
fission studies is its ability to describe states at very large deformation.
Two different types of properties control the general
features of fission dynamics: on the one hand the surface and surface-symmetry
energy coefficients that determine the average resistance of
the nucleus against deformation \cite{(Nik11a),(Jod16a)}, and on the other hand the evolution
of shell structure that generates the minima and maxima associated
with the multi-humped structure of the deformation energy landscape
\cite{(Bra72a)}.

There is some direct information about the excitation energy of
highly-deformed states that is available and that can be used to
inform the parameter fit. On the one hand, there are barrier heights data
\cite{(RIPL),(RIPL3)}, which have to be interpreted with some caution
as in one way or the other the available values were obtained via
intermediate models \cite{(RIPL)}. On the other hand, there are also
measured excitation energies of some fission isomers \cite{(Sin02)}.
For a very limited number of fission isomers there is also
information about their quadrupole deformation from $E2$ transition
moments \cite{(Met80),(Thi02)}, and some information about their shell
structure can be obtained from the quantum numbers of bandheads.
Additional data on such states would clearly be of great help for
fine-tuning  the nuclear EDF.

To date, the EDFs most commonly used in fission studies have been adjusted to
fission isomer excitation energies \cite{(Kor12)} or fission barriers \cite{(Bar82c),(Ber91),(Gor07)},
with the exception of the relativistic functionals PC-PK1 \cite{(Zha10)} and DD-PC1 \cite{(Nik08)} that
combine information on deformed heavy nuclei and the nuclear matter equation of state.
Some authors suggest paying more attention to the nucleus-nucleus
interaction between pre-fragments  near scission \cite{Adamian16}.

A technical issue that needs to be addressed
is that many parameterisations
of the nuclear EDF exhibit so-called finite-size instabilities,
meaning that homogeneous infinite nuclear matter is unstable against
a transition to an inhomogeneous phase that is either polarised in
spin or isospin or both, see \citeasnoun{(Pas15a)} for a review.
Finite-size instabilities can be
triggered by gradient terms in the EDF, but also by
finite-range terms in non-local EDFs \cite{(Mar19a),(Gon20)}.
Many parameterisations of the Skyrme EDF exhibit such instability in
one or the other spin channel, which becomes an issue when
working with time-reversal breaking configurations or when
calculating certain RPA modes. Such instabilities are
also sometimes found in isovector channels of some Skyrme and Gogny
parameterisations.  All of
these instabilities can be efficiently and unambiguously detected
with linear-response calculations of infinite nuclear matter. Such test can be easily incorporated into fit
protocols, as already done for the UNEDF2 \cite{(Kor14)} and SLy5sX
\cite{(Jod16a)} Skyrme parameterisations.

Irrespective of the choices that will be ultimately made for the form
of the EDF and the protocol for the adjustment of parameters,  it is
clearly desirable to have just one or a few standard
EDFs for fission studies that are used by as many groups
as possible in order to eliminate possible dependencies upon the
parameterisation when comparing results obtained with
different approaches to treat the many-body problem.
For TDDFT treatments, it is also important that EDFs are fitted
without the centre-of-mass corrections \cite{(Goe83),(Ska06),(Kim97a),Simenel2012,(Kor12)}.

\subsection{Uncertainty quantification}
\label{subsec:stats}

As discussed in the previous section,  nuclear density
functionals have to be calibrated to experimental data. This empirical wisdom is built into the
quality measure $\chi^2(\vec{p})$ which is a scalar function of the $N_p$
model parameters $\vec{p}$. The common use of $\chi^2$ is to
deduce the optimal parameterisation $\vec{p}_0$ by minimising $\chi^2$.

Systematic uncertainties can be revealed by comparing predictions of
different models; for fission applications, see
\citeasnoun{(Kor12),(Agb17)}. In the context of statistical
uncertainties related to model parameters, much  information can be
unraveled by employing $\chi^2$ in connection with the  tools of
statistical analysis
\cite{(Dob14),(McD15),(Sch15a),(Sch15c),Niksic15,(Rei18)}.

Computing the probability distribution of the parameters $\vec{p}$ rather than
a single point gives immediately access to two important new pieces of
information, the uncertainty of a predicted observable,
and the
correlation between two observables.
Uncertainties are important to control the quality of a
prediction. This is mandatory when using the results in further
calculations as done, e.g., in nuclear astrophysics, and it is an
extremely useful indicator for model development because it reveals
deficiencies of parameterisations. Correlations add another
world of information. They allow a sensitivity analysis to check the
impact of a certain model parameter on an observable
\cite{(Kor10c),(Kor12)} and they indicate the information content of a new
observable as compared to previous ones \cite{(Rei10a),(McD15)}.
In the context of the present report, it is particularly interesting
to apply correlation analysis for the very different observables
discussed here, e.g., relating fusion cross sections and fission
properties.

There is still more potential in statistical analysis of DFT. So far, the
evaluation of uncertainties and correlations has mostly been based on a Taylor
expansion of the $\chi^2$ and of observables around the optimal
parameterisation $\vec{p}_0$. This runs easily beyond validity,
particularly for fission properties \cite{(Hig15)}. Thus one needs to evaluate the
integrals $\int d^{N_p}p ...$ in detail which grows quickly infeasible.
Here one can take advantage of modern  techniques of supervised learning.
Employing the posterior probability distribution computed with emulators,
one can propagate theoretical statistical uncertainties in predictions of
various computed quantities, including binding energies and PESs \cite{(McD15),(Neu18),Neu20,(Las20)}.
One can teach the emulators to improve the
predictions of selected observables in a given region of the nuclear
chart by one order of magnitude at practically no extra cost. This is
particularly desirable if the output of nuclear DFT calculations is used
as input in other chains of calculations as, e.g., in nuclear astrophysics simulations.

To estimate  uncertainties, both systematic and statistical, uniform
model mixing \cite{(Erl12d),(Agb17)} can already provide a very
valuable information. More advanced techniques involve Bayesian model
averaging   \cite{(Neu19),Neu20}, which allows to maximise the ``collective
wisdom" of  relevant models by providing the best prediction rooted
in the most current experimental information. This will be an
important part of future collaborative projects in fission theory.


\section{Fission fragments}
\label{sec:fragments}

In the last stage of the fission process, the nucleus descends towards scission
where it divides into nascent fragments,  which then
de-excite, see {\Fig}~\ref{fig:fission_physics}. The following sections
are devoted to fission fragments and the related fission observables.


\subsection{Scission}
\label{subsec:scission}

The scission event is arguably one of the least understood processes
in fission, although some experimental information on scission configurations
have just became available \cite{Ramos2020}. In a mean-field picture it marks the transition from the
final state of the elongated fissioning nucleus to the initial state
of the two separated nascent fission fragments. As discussed later in
{\Sec}~\ref{subsec:entanglement}, there are good reasons to think that the nascent
fragments are entangled at, or immediately after, scission but it is
not clear whether this entanglement persists to the stage of the fission fragments or quantum
decoherence takes place. Furthermore, the characteristics of these
fragments such as their charge, mass, energy, angular momentum,
parity, level density, etc., are crucial ingredients in determining
the properties of the neutron and $\gamma$ spectra, as well as the
$\beta$-decay chains, see {\Sec}~\ref{subsec:spectrum}.

Before microscopic time-dependent descriptions of fission dynamics
became available, the scission event was most often treated with
\emph{ad hoc} assumptions, ignoring any role for dynamics. At one extreme, scission
is assumed to transform
the system into a statistical ensemble of two nuclei having their
surfaces separated. In the framework of
DFT theory, various criteria were introduced to define scission based
on the HFB solution for the fissioning nucleus. The simplest ones
define a threshold value for the density between
the two pre-fragments or the expectation value of the  neck
moment. The main problem with such  schemes, however, is that
both the intrinsic energy of each pre-fragment and their relative
interaction energy are extremely poorly described: before the
separation of the pre-fragments, both the nuclear and Coulomb interaction
energies are vastly overestimated because of the large overlap between
the pre-fragments; when the primary fragments are well separated, the
minimisation principle underpinning the HFB equation leads to the two
fragments in their ground state. Such dramatic simplifications can be
mitigated by performing unitary transformations on the total wave
function of the fissioning system which, while leaving the whole
system invariant, can be designed to minimise the interaction
(or equivalently, maximise the localisation) of the pre-fragments
\cite{(You11)}. In spite of the development of such  techniques, it is clear
that explicitly non-adiabatic, time-dependent methods provide a much
better handle on scission -- at least when it comes to defining the
initial conditions of the nascent fragments.


\subsection{Fission fragment yields}
\label{subsec:yields}

Three main methods are used to determine the yields of
different fission fragments. Scission-point models assume
a statistical distribution of probability among a set of
scission configurations of the
nucleus \cite{fong1953,wilkins1976}, see
\citeasnoun{(Lem19)} and \citeasnoun{(Pas19)} for recent
realisations.
These models
require the definition of an ensemble of scission configurations that
can either be determined by constrained mean-field
calculations or from an analytical parameterisation of the
shapes of the di-nuclear system. Each of these nuclear configurations is then
populated according to a Boltzmann distribution, with the temperature
defined in accordance with the initial energy of the system.
Since it is computationally effective, this method is used in systematic studies
or to investigate the
evolution of yields with the excitation energy of the compound
nucleus. However, the choice of the ensemble of scission
configurations remains arbitrary and may influence the resulting
yields. Moreover, an explicit use of temperature for a non-adiabatic
and time-dependent process is not really well justified.

To some degree, also the total fragment kinetic energy may be estimated.
In particular, some models have sought to predict those quantities
exclusively on the basis of the scission configurations \cite{lemaitre2015a},
but their predictive power has been limited due to the importance
of the collective path taken prior to scission.
Indeed, experience with  a diffusive transport model
\cite{(Ran11c)} has shown that not only the scission hypersurface
but the global topography of the PES may have a qualitative influence
on the outcome.

To avoid the assumption of statistical equilibrium at scission, one
possibility is to describe the evolution of the compound
nucleus from some initial state at lower deformation  up to the configurations
close to scission. In this approach, one defines an
equation of motion for a few collective coordinates associated with a
parameterised shape of the nuclear surface. Assuming that these
collective degrees of freedom interact with a thermal bath of
intrinsic excitations, leads to the Langevin equations in the deformation space; see {\Sec}~\ref{dissdyn}.

These transport equations have been solved in multi-dimensional spaces both in MM and hybrid-DFT frameworks
either directly \cite{(Miy19),sierk2017,ishizuka2017,usang2019,(Sad16a),Sadhukhan2017,matheson2019} or in the strongly damped
(Smoluchowski) limit \cite{(Ran11b),(Ran11c),(War17)}.
The result is the probability
of populating different nuclear configurations close to scission
and it is then straightforward to determine the resulting mass asymmetry.

Most often these treatments have concentrated on the mass number,
assuming that the proton-to-neutron ratio remains constant,
but recent progress has been made
\cite{(Mol14),(Mol15a),MollerEPJA53,(Sad16a),Sadhukhan2017,matheson2019}
towards including also the isospin degree of freedom,
thus making it possible to determine both the mass and the charge yields.
The transport framework takes into account the dissipative effects of
the collective dynamics and may even account for the emission of neutrons
in the course of the evolution \cite{(Esl18)}.
One general limitation is that the Langevin treatment
is restricted to the classically allowed region of the collective space,
so it cannot treat tunnelling.

As mentioned in {\Sec}~\ref{secGCM}, an alternative approach to
computing fission fragment yields within a quantal description
is the
TDGCM \cite{(Reg19a),(Zha19),younes2019,(Ver20)}.
Here, a major difficulty is the
determination of a proper manifold of states which usually consists
of an ensemble of quadrupole and octupole constrained
HFB solutions.
While this description of quantum collective dynamics can treat tunnelling,
it fails to include diabatic aspects of the dynamics close to scission.
Let us also mention recent
TDHF \cite{SimenelUmar2014,(God15a),Goddard2016},
TDHF+BCS \cite{(Sca15b),scamps2018} and TDHFB \cite{(Bul19a)} studies
of fission.
In these cases, the initial configurations for the time-dependent
calculations are generated by constrained calculations at some
elongation beyond the outer turning point.
These methods are well suited to investigate the role of shell
effects at scission \cite{scamps2018,ScampsSimenel2019},
and thus provide valuable guidance to more phenomenological
models like the scission-point models discussed earlier in this
section.
As mentioned earlier in {\Sec}~\ref{subsubsec:dynamics},
time-dependent theories will be challenged to reproduce the tails of
the yield distribution, due to non-Newtonian Langevin trajectories,
unless a mechanism equivalent to the random force of the Langevin
equation is included \cite{aritomo2014,Sadhukhan2017}.
Moreover, the present formulation does not allow for the treatment of quantum tunnelling.

All these state-of-the-art methods have their own strengths and weaknesses. Yet,
they all rely on determining the probabilities to populate a set of scission
configurations.
A common feature in all these approaches is that the fragment yields
are computed at scission, where the two nascent fragments still interact
through the nuclear force, see {\Sec}~\ref{subsec:scission}.
As a result, estimates of particle number with projection methods, for example,
become extremely sensitive to the condition that define scission
configurations.
Other observables
such as energy sharing between nascent fragments, may not be relevant
at this stage of the fission process. Methods should be developed to
determine the yields of observables further away from scission.

Current methods have been mostly focused on the yields associated
with the mass, charge, and sometimes TKE of the fragments.
To go beyond this simple picture, the challenge is to extend the space of
configurations in the fission channel in order to be able to make
quantitative predictions of correlated yields for these three
observables, and eventually additional ones. The new observables of interest are
typically the angular momentum and parity of the nascent fragments.

Finally, the prediction of fission yields is essential for a correct description
of $r$-process nucleosynthesis and superheavy elements.
It would, therefore, be important to carry out systematic large-scale  calculations of
fission yields in regions of the nuclear chart far from the valley of stability.
While such large scale calculations present a serious challenge for computationally
intensive models of fission dynamics, some recent progress in this direction has been reported in
\citeasnoun{(Giu18),(Lem18),giuliani2019a,RGRobledo2020}.


\subsection{Number of particles in fission fragments}
\label{subsec:particles}

The estimation of $Y(Z,A)$ is usually based on the
assumption that the probability density of the mass and charge of the
fragments associated with a Bogoliubov wave function is Gaussian.
 However, in order to describe the odd-even staggering seen in
charge distributions more refined methods are required.

In the MM DDD approach, the odd-even effect in fission yields can be attributed
to the pair-breaking effect \cite{Mir14}.
In order to assign a particle number to a pre-fragment in the vicinity of the scission
configuration, a condition has been introduced in \citeasnoun{Mirea11} based on the
position of the neck.
A similar approach to particle number identification  was proposed in the DFT approach of \citeasnoun{(You11)} using
the unitary transformations on the total wave
function aiming at maximising the localisation of pre-fragments.

Even earlier,  a method has been proposed in \citeasnoun{Sim10} to estimate
the exact probability distribution of mass and charge in a nascent fragment  created in
microscopic models by introducing
the particle-number projection for fragments. This method has been applied to
determine the transfer probabilities of nucleons during collision reactions and then
generalised to superfluid system \cite{(Sca13)} with the use
of the Pfaffian method \cite{robledo2009}, and applied to
fission \cite{(Sca15b)} in TDHF+BCS, thus showing that the odd-even effects
can be described with the mean-field dynamics.
As discussed in {\Sec}~\ref{subsubsec:diffusive}, these distributions are affected by the
lack of one-body fluctuations and correlations (e.g., between mass
and charge distributions). As shown in \citeasnoun{Simenel2011,Wil18,Godbey2020},
the latter can be recovered to some extend
for symmetric systems using the TDRPA \cite{(Bal84a)}.

It will be interesting to couple this approach with
configuration-mixing methods and semi-classical descriptions of the
fission process. One should also go beyond the  approximation of identical
occupation of time-reversed canonical HFB states assumed in
\citeasnoun{(Ver19)} to see whether the proper blocking of
one-quasiparticle states in odd-$A$ nascent fragments,
associated with breaking of time reversal symmetry, is important for
the description of odd-even staggering of fission yields.

%

\subsection{Energy sharing}
\label{subsec:sharing}

Most of the energy released in fission appears in the form of
TKE of the fission fragments. Hence, a direct inverse
correlation exists between TKE and their total excitation energy
available for prompt neutron and  gamma emission.
Moreover, the distribution of TKE directly influences the prompt
neutron multiplicity distribution, which has been
measured in a few cases and is important in transport simulations of
selected classes of integral experiments.

Once the nascent fragments are separated at scission, the Coulomb repulsion
is transformed into kinetic energy. As indicated in {\Sec}~\ref{subsubsec:ecoll},
 however, different models predict different values
for the collective kinetic energy at scission. It is
typically a few MeV in TDDFT
and ranging from zero to 20 MeV in various transport treatments.
From a theoretical point of view a tolerance of 20 MeV, representing about
10-15\% relative uncertainty, might be deemed acceptable.
However, a change of TKE by that much would significantly change
the multiplicity of evaporated neutrons (by about two)
and it is therefore an important challenge to fission theory
to improve on the calculation of TKE.

The available total excitation energy in fission fragments can be
 calculated from the energy balance in a fission event, knowing
the masses of the fissioning system and of the fission fragments,
once those are determined via a chosen theoretical model, or
extracted from systematics of experimental data. For any model that
does not fully separate the fission fragments, the extraction of the
energy sharing will be subject to large uncertainties, as energy can
flow from one pre-fragment to the other through the neck, and in close
proximity the nascent fragments exchange energy via Coulomb interactions.
Moreover, the nascent fragments are generally distorted relative
to their equilibrium shapes and the associated distortion energy
will be converted to additional primary fragment excitation energy,
thus affecting the resulting energy sharing.

Guidance on excitation energy partitioning is necessary for
simulating neutron and photon emissions when the total excitation
energy in the fissioning system increases (as in the case of
fission induced by fast neutrons). The only indirect
observable related to the excitation-energy sharing is the average
number of neutrons per fission event as a function of mass, but the
data beyond thermal neutron-induced fission and spontaneous fission
reactions are scarce. The results of average
neutron multiplicity measurements  as a function of mass for significantly
different excitation energies in the fissioning systems,  suggest that with increasing energy in the fissioning (actinide)
system, most of the extra energy is deposited in the heavy fragments
\cite{Muller:1984,Navqi:1986}.

It has to be emphasised that  the process of energy sharing poses an
interesting and nontrivial question: does the energy sharing occur
in the condition of thermal equilibrium that has developed between
nascent fragments? In this case, the details of the process of neck
formation and subsequent fracture would be of secondary importance,
and only the density of states associated with each of the nascent
fragments would play a role. On the other hand, if equilibrium is not
reached during the saddle-to-scission evolution the details of the
splitting process will be crucial. This issue is still not resolved.
Namely,  the TDDFT method, which has recently been used to
parameterise the energy dependence of the excitation-energy sharing,
predicts neutron multiplicities as a function of mass in
agreement with experimental observations
\cite{(Bul19a),Bulgac:2019dyn}.
At the same time, an approach that
models the excitation-energy sharing statistically on the basis of
the microscopic level densities within a Brownian shape evolution
framework, was also able to reproduce the experimental trend
\cite{(Alb20)}.


\subsection{Quantum entanglement}
\label{subsec:entanglement}

Spontaneous fission of even-even nuclei is a process by which
a $0^+$ quantum system decays into two excited nuclei
which eventually, after prompt neutron and photon emission,
turn into two product nuclei in their ground states,
moving apart with opposite momenta.
In this sense, the process is analogous to the emission of
two electrons from a singlet state,
with the additional complication that in fission,
neutrons and gamma rays are emitted at or beyond the scission point.

The fission process conserves quantum numbers and, therefore, those that
characterise the initial state,
such as particle number  \cite{(Bul17)} and angular
momentum, must be shared among all particles and quanta in the exit
channel. For example, neglecting neutron emission, the final
state would be a superposition of states of the two fragments with
numbers of protons and neutrons in one fragment complementing those
in the other fragment, so that they add up to the number of protons and
neutrons of the initial fissioning nucleus. The particle numbers of the
fragments are therefore entangled, and a measurement in one
fragment collapses the information about the particle numbers in the
other fragment.

The same is true for the measurement of $\gamma$ rays, which are
characteristic of a given nucleus and thus uniquely define the
other fragment. Summing up the angular momenta of gamma
rays emitted from one primary fragment collapses the information about the
angular momentum of the other fragment in the exit channel. In the same way
the angular-momentum polarisations of the two fragments are also entangled.
The real question is whether these effects are ultimately important for
experiment? Can they be observed at all? When and how does
decoherence of this entanglement occur? Fission fragments may
represent a unique opportunity to explore quantum entanglement
of mesoscopic systems, that is, they can be the closest
realisation of the Schr\"odinger cat phenomenon \cite{(Dob19a)}.


\subsection{Quantum numbers}
\label{subsec:quantum}

An essential ingredient in
the microscopic description of fission is the PES, constructed in
an intrinsic frame and including pairing correlations in the BCS
or HFB approximations, see {\Sec}~\ref{subsec:pes}.  Both of these break symmetries of the
underlying Hamiltonian, and their restoration yields additional
contributions to the PES; see discussion in {\Sec}~\ref{subsec:params}.

The restoration of particle number has a relatively small effect on the PESs
in actinide nuclei, which are far from closed shells \cite{Bernard2019}. However, it may
have a significant impact on the ATDHFB or GCM inertias that determine the
collective Hamiltonians; see {\Sec}~\ref{subsec:lubricant}.
In particular, extensive studies are needed in two specific areas.
First, one should consistently calculate the PES
for symmetry-restored wave functions in
the case of particle number projection, possibly by including variation
after projection to determine the intrinsic states. Second, and perhaps
most important, is the development of a consistent theory of collective
inertias for the symmetry-restored wave functions.
Recent developments in the description and manipulation of particle
number projected HFB states (called Antisymmetrized Germinal Power in the
quantum chemistry literature) \cite{Dukelsky2019}
could potentially lead to a particle-number-projected ATDHFB theory.

The angular momenta of fragments is an important
element that determines neutron yields and other decay properties \cite{Wilhelmy1972}.
Given a mean-field description of the nucleus and the knowledge of
its quasiparticle excitation energies, current theoretical tools can
be used to calculate the angular momentum content
of the fragments.  There are two components to the angular momentum of
the newly formed fragments: non-collective and collective.

The non-collective angular momentum is carried by the quasiparticles in the pre-scission configuration that  are transferred
to the post-scission nascent fragments under diabatic conditions.  They will
end up in one or the other nascent fragment, depending on the evolution of the corresponding orbitals with
elongation; see \citeasnoun{(Ber19c)} for an example of this transition.
Their angular momentum is conserved, allowing
one to estimate its contribution to that of the nascent fragment.
The  collective angular momentum
arises because of the deformation of the compound nucleus.
This component can be calculated by well-known projection techniques used to compute
rotational bands in deformed nuclei.
The only difference in the case of the fissioning system is that scission
also affects the angular momentum of the system with respect to
the orientation of the fission axis.
The collective contribution to the angular momentum of the nascent fragments
about the fission axis vanishes. As a result, the distribution of
gamma radiation in the subsequent cascade will be anisotropic with
respect to this axis \cite{Bertsch2019a}.
In fact, this anisotropy has been observed in spontaneous fission,
and a systematic measurement would provide an invaluable test of our overall
understanding of the dynamics at the scission point.

There can be additional angular momentum generated as the pre-fragments
separate, due to higher multipole components of the Coulomb
field between them, see \citeasnoun{(Ber19d)}. It is, in fact, straightforward
to calculate the effect of the electric quadrupole field on the post-scission nascent fragments,
given their deformations and their initial separations.  It would
therefore be useful to have this information available when reporting
fission calculations going through the scission point.


\subsection{Fragment de-excitation}
\label{subsec:spectrum}

Nascent fragments emerge with significant excitation energies and then primary fragments cool down
via various decay modes resulting in particle emission
(neutrons and photons from prompt and delayed emission
as well as electrons and antineutrinos from $\beta$ decays).
In current phenomenological approaches, the neutron emission proceeds after
the nascent fragments have fully accelerated becoming primary fragments, whereupon they
are treated as compound nuclei that de-excite via particle emissions.

In order to carry out simulations of those decay chains, it is necessary
to know the initial states of the primary fragments, in particular their
initial excitation energy, angular momentum, and parity. On the other hand,
experimental information regarding the fission fragments can only be
obtained after neutron emission. Hence, few experimental data can
inform phenomenological models, and the microscopic models can
play an important role in providing the necessary input for a large
range of reactions. The angular momentum of the emerging fission fragments is an
important quantity that sets the competition between the neutron and
$\gamma$ emission, and it has an important influence on a variety of
photon observables, from prompt fission $\gamma$ multiplicity to the prompt
fission spectrum and correlations between emitted photons.
To a lesser degree,
it can also influence the delayed neutron and $\gamma$ properties.

To reduce uncertainties, the angular momentum
properties of the fission fragments should be investigated in a
framework that allows the total separation of the nascent fragments.
It has been demonstrated in TDDFT that scission is followed by
a relaxation period  in which the nascent fragments transition to a
deformation of primary fragments that is close to the ground-state deformation
\cite{(Bul19a)}, thereby increasing the energy available for emission.

Both
$\gamma$ and $\beta$ decays of the primary fragments can be described
using the QRPA, thus defining a consistent framework
both for the entrance and exit channels, see {\Sec}~\ref{sec:entrance}.
Because $\beta$-decay half lives are long compared to those of $\gamma$ decays,
they generally can be assumed to occur from the fragment ground state,
thus making a finite-temperature description of $\beta$ decay unnecessary.
But it is important to take into account that the $\beta$ decays
may generally populate excited states in the daughter fragment,
which would then undergo their own emission chain
before a subsequent $\beta$ decay could occur.
Consequently, $\gamma$ decay should be investigated for each primary fragment
as a prompt phenomenon (in principle in competition with but usually
after neutron evaporation). $\beta$ decay to the resulting
ground-state fission products should also be investigated, including
forbidden transitions of particular importance for  neutrino studies.


\section{Computational strategy}
\label{sec:computation}

Microscopic modelling of nuclear fission is an example of a
computational grand challenge
\cite{young2009,bishop2009,exascale2016}. One reason is the
complexity of the problems, whose solutions require advanced notions
of linear algebra, group theory, analysis, computer science, etc.
Another reason is the sheer amount of computing needed. While a
single, static HFB calculation may take between a few minutes to a
few days to converge on a single CPU, depending on the functional,
the number of broken symmetries and the types of constraints, up to
dozens of millions of such calculations would need to be performed to
tackle some of the problems discussed in this document (large scale
potential energy surfaces, functional optimisations, time evolution,
action minimisation, symmetry restoration). In this section, we
discuss the various computational strategies that are currently
available or should be explored in the future.


\subsection{Computer codes for fission}
\label{subsec:codes}

There is a broad consensus that fission is not a problem that can be
handled by a single code. Instead, the community should think of an
ecosystem of different frameworks addressing different facets of the
problem, for example, static versus time-dependent calculations.
Since fission calculations are almost always characterised by the
need to compute and manipulate very deformed configurations in heavy
nuclei, this imposes specific requirements about the codes. In this
section, we review some of the existing software and identify current
gaps.


\subsubsection{Brief review of available codes}
\label{subsubsec:available}

A number of computer codes are available for modelling various aspects of
fission dynamics. The most computationally intensive ones are those
used for the calculation of HFB configurations and related quantities
(inertias, energy overlaps, time-dependent evolution, etc.). Two basic implementations
of HFB/TDHFB solvers differ in the representation of the single-particle wave functions:

\begin{description}[align=left]
\item [Space Discretisation] This family of codes employs a finite volume,
either in coordinate space or momentum space, that is
discretised using a suitable mesh of points.
Many different choices can be found in the literature, from the Lagrange mesh
based on orthogonal polynomials \cite{(Bay15)}, to B-spline-based \cite{Blazkiewicz2005,(Pei08)},
and adaptive wavelet-based methods \cite{(Pei14)}. Among the codes relevant for fission
we mention: Sky3D \cite{(Mar14),Afibuzzaman18,(Sch18)},
and EV8 \cite{(Rys15)} (publicly available) as well as
HFB-AX \cite{(Pei08)}, MADNESS-HFB \cite{(Pei14)},
MOCCa \cite{(Rys16)}, HFB-2D-LATTICE \cite{Teran2003,Blazkiewicz2005},
Skyax \cite{(Rei20a)},
and the code of \citeasnoun{(Jin17)}.

\item [Basis Expansion] This family of codes is based on an expansion of
single-nucleon wave functions in a finite set of suitable basis functions.
Most often this is a basis of the harmonic oscillator (HO) eigenfunctions. There
have also been attempts to use the transformed HO basis
states \cite{(Sto03)}, or a two-centre HO basis for improving the
description of elongated shapes \cite{(Dub98)}. Some of the principal
codes that use this representation for fission modelling are:
HFODD \cite{(Sch17b)} and HFBTHO \cite{(Per17)} (both publicly
available,) as well as HFBaxial \cite{Robledocode,(Rob18)}, HFBTri \cite{(Rob18)},
and HFB3 \cite{(Has13)}. The
code HFB3 utilises
a basis expansion for two Cartesian directions while employing a
Lagrange mesh in coordinate space in the third. Two codes based on relativistic EDFs
have recently been used in calculations of self-consistent mean-field
configurations as input for modelling fission dynamics: DIRHB \cite{(Niksic2014)} and
MDC-RMF \cite{(Lu2014)}.
\end{description}

The use of different
representations determines the applicability of any given code.
For instance, all numerical schemes can efficiently deal with zero-range
(Skyrme-like) effective interactions; however, apart from
implementations in spherical symmetry \cite{(Ben20a)}, mesh-based discretisation
schemes have not been able to employ finite-range effective forces
so far. On the other hand, basis-expansion methods have a long
history of calculations with (among others) various Gogny
interactions \cite{(Rob18)}, Coulomb \cite{(Dob96b)}, and Yukawa
\cite{(Dob09g)} forces, by expanding the interaction into
Gaussian form factors. Extensions to more general finite-range
interactions have also been proposed \cite{(Par13a)}.

For the time-dependent calculations of fission dynamics, only mesh-based methods
have been used so far. They satisfy the demands of extreme
deformations encountered at scission, and also allow the description
of the nascent fragments beyond scission.
While techniques exist to optimise the choice of basis states, mesh-based
calculations have the advantage that their numerical precision is essentially
independent of the nuclear deformation \cite{(Rys15a)}.

A major aspect of the difference between the two numerical schemes is
the discretisation of the continuum. The first consequence of this is
the treatment of pairing correlations since mesh- and basis-based methods
give very different descriptions of positive-energy single-particle states.
On the one hand, the
coordinate representation takes correctly into account asymptotic
properties of single-particle wave functions, but it becomes impractical and
time-consuming when dense single-particle states extending to high energies
are included. This is required for a full HFB description of the
coupling between quasiparticle and quasihole states \cite{(Dob84a)}. On
the other hand, basis-expansion methods can easily manage states
of arbitrarily high energy, but fail to reproduce the spatial asymptotics
of wave functions, and are thus not appropriate for the description of
weakly-bound systems. In view of the importance of pairing for
fission applications, see  {\Sec}~\ref{subsec:lubricant}, this
situation is not satisfactory. One should note, however, that the practical
importance of asymptotic properties or high-energy states in the continuum
has not yet been evaluated for fission.

Another aspect is the treatment of finite-temperature calculations.
When the nucleus is heated  the Fermi surface becomes more diffuse
and the statistical mixture includes contributions from quasi-bound
and unbound single-particle states. While preliminary studies have been
reported \cite{(Bon84),Zhu2014,Schuetrumpf16T}, the correct treatment of this degree of
freedom is an open problem, see {\Sec}~\ref{subsec:temperature}.
Nevertheless, as in the case of pairing correlations, we can already
foresee that the applicability and performance of mesh-based and basis-based
approaches will differ significantly.

\begin{table}[!htb]
\caption{\label{tab:codes]}
Summary of deformed HFB and TDHFB solvers. RHB stands for relativistic HFB.
References to codes: [1] \protect\cite{(Rei20a)},
[2]  \protect\cite{(Mar14),Afibuzzaman18,(Sch18)},
[3]  \protect\cite{(Rys15)},
[4]  \protect\cite{(Pei08)},
[5]  \protect\cite{(Pei14)},
[6]  \protect\cite{(Rys16)},
[7]  \protect\cite{(Jin17)},
[8]  \protect\cite{(Kim97a),Simenel2011,(Sca13)},
[9]  \protect\cite{sekizawa2016,Wil18},
[10] \protect\cite{UmarOberacker2005},
[11] \protect\cite{(Sch17b)},
[12] \protect\cite{(Per17)},
[13] \protect\cite{Robledocode},
[14] \protect\cite{(Rob18)},
[15] \protect\cite{(Has13)},
[16] \protect\cite{(Niksic2014)},
[17] \protect\cite{(Lu2014)}.
}
\begin{indented}
\item[]\begin{tabular}{lcll}
\br
\multicolumn{4}{c}{\bf Coordinate space representation}\\
\mr
SkyAx       & [1] & 2D axial                & static CHF+BCS      \\ 
Sky3D       & [2] & 3D Cartesian            & CHF+BCS/TDHF        \\ 
EV8         & [3] & 3D Cartesian            & static CHF+BCS      \\ 
HFB-AX      & [4] & 2D axial, B-splines     & static CHFB         \\ 
MADNESS-HFB & [5] & 3D wavelets             & static HFB          \\ 
MOCCa       & [6] & 3D Cartesian            & static CHFB         \\ 
LISE        & [7] & 3D Cartesian            & HFB/TDHFB           \\ 
TDHF3D      & [8] & 3D Cartesian            & TDHF/TDRPA/TDHF+BCS \\
3DTDHF      & [9] & 3D Cartesian            & TDHF/TDRPA          \\
VU-TDHF3D   & [10]& 3D Cartesian            & TDHF (density constraint) \\
\mr
\multicolumn{4}{c}{\bf Basis expansion}\\
\mr
HFODD       & [11]& 3D HO                   & static CHFB         \\ 
HFBTHO      & [12]& 2D axial HO             & static CHFB         \\ 
HFBaxial    & [13]& 2D axial HO             & static CHFB         \\ 
HFBTri      & [14]& 3D HO                   & static CHFB         \\ 
HFB3        & [15]& 2D HO $\otimes$ 1D mesh & (TD)HFB             \\ 
DIRHB       & [16]& 3D HO                   & static C RHB        \\
MDC-RMF     & [17]& 2D axial HO             & static C RHB        \\
\br
\end{tabular}
\end{indented}
\end{table}

Table \ref{tab:codes]} summarises the available codes
for modelling deformed nuclei, their collective properties and fission dynamics.
A few additional comments are in order: in the code Sky3D all spatial derivatives are evaluated using the finite Fourier transform method;
the code of \citeasnoun{(Has13)} uses a Lagrangian
coordinate-space grid in the direction of the axial-symmetry axis;
the code of \citeasnoun{(Jin17)} uses a complete basis of single-particle states
in the solution of the TDHFB; code SkyAx (HFODD) can implement
constraints on axial monopole, quadrupole, octupole, and hexadecapole
deformations (non-axial deformations up to multipolarity $\lambda=9$) separately.


\subsubsection{Development of new capabilities}
\label{subsubsec:new}

Even if the codes listed above offer a high level of flexibility and great potentiality, we recommend
the development of the following new computing capabilities.
\begin{description}[align=left]
\item[Adaptive meshes] The specific issues of fission dynamics require
       that nuclear wave functions are computed also in regions where the
       nuclear density vanishes. For mesh-based
       implementations, uniformly spaced grids thus include
       discretised continuum represented by a large
       number of single-particle states, even at fairly low
       energies. For this reason, we recommend the development of a new-generation
      of  mesh-based codes that will utilise non-uniform meshes, with
       lattice points concentrated in space regions where
       nuclear densities are non-negligible. This method
       would require self-consistent redefinitions of meshes
       depending on relative distances, deformations, and
       relative orientations of nascent fragments.

\item[Adaptive bases] By definition, implementations that utilise two-centre
       HO bases describe only regions of space where
       densities are sufficiently different from zero. However, they
       require adaptive methods to self-consistently define bases
       corresponding to relative distances, deformations, and
       relative orientations of nascent fragments, as proposed in
       \citeasnoun{(Dob19a)}. We recommend the development of
       the corresponding HO-basis codes. Another direction is to use
       the multi-resolution  techniques with a multi-wavelet basis
       as in \citeasnoun{(Pei14)}.

\item[General symmetry breaking] The concept of spontaneous symmetry
       breaking is crucial for a mean-field description of atomic nuclei,
       but it has not been exploited to its full capabilities yet.
       Even the most ambitious fission studies to date consider only
       configurations that still maintain certain self-consistent symmetries. In
       particular, time-reversal breaking configurations, needed for
       the description for odd and odd-odd nuclei as well as
       high-spin physics and multi-quasiparticle configurations, are not included
       in most available computer codes. To the best of
       our knowledge, HFODD and Sky3D are the only publicly available codes
       that include the degrees of freedom necessary for this type of
       studies.

\item[GCM with time-odd momenta] Along the same line, an extension of
       current GCM codes to include time-odd collective momenta
       presents an interesting challenge. In addition to the
       implementation of time-reversal symmetry breaking,
       and the development of the actual GCM, this would also
       require the capability to consistently construct and constrain
       the relevant conjugate momentum operators.

\item[Overlaps and Kernels]
       Both GCM and projection methods require
       multi-reference calculations, that is, determination of
       overlaps and matrix elements between different paired or
       unpaired product states. In addition to possible problems related
       to the density-dependence of the interaction
       \cite{(Dob07),(Lac09),(Ben09c),(Dug09),(Rob10b),(She19a)},
       such calculations
       always require a higher degree of symmetry breaking than
       those for the corresponding single-reference implementations.
       For example, restoration of the particle symmetry or 3D
       rotational symmetry implies the time-reversal or simplex
       symmetry breaking, respectively, even if the single-reference
       states that are subjected to projection conserve these
       symmetries. Therefore, it is recommended that new
       codes are initially developed with a maximum degree of
       symmetry-breaking capabilities, and then accelerated by
       implementing conserved symmetries in single-reference
       calculations. This will ensure that they are automatically
       portable to multi-reference frameworks.

\item[Matrix Elements]
      To implement $K$-matrix reaction theory for fission rates, as discussed in
      {\Sec}~\ref{subsec:reaction}, the suite of computer programs should be
      augmented with routines that calculate effective Hamiltonian matrix
      elements between arbitrary CHF and CHFB configurations.  Also,
      the calculation of decay widths require including
      momentum operators in the set of constraining fields in the CHF and CHFB
      codes.

\item[TDDFT, TDGCM, ATDDFT, and QRPA codes]
       We recommend building new-generation codes for fission dynamics by
       directly implementing the capabilities
       of the TDDFT, TDGCM, ATDDFT, and QRPA methods. For example,
       implementations built on HO bases, proposed in \citeasnoun{(Dob19a)}, can be ported
       to the time-dependent adaptive bases for TDDFT,
       or to the iterative solutions of the ATDDFT and QRPA methods.

\item[Mesh-based codes for non-local EDFs]
       Since it is unlikely that higher accuracy of the calculated
       nuclear observables can be obtained using local EDFs \cite{(Kor14)},
       current developments are focused on new-generation
       non-local EDFs. As discussed above, codes based on the HO basis
       are capable of treating such functionals fairly efficiently. It
       is, therefore, of paramount importance to develop algorithms
       for implementing the same capabilities in
       mesh-based codes. This is certainly a far-reaching goal;
       presently with no clear ideas on the direction to take.
       Nevertheless, we recommend that, because of its fundamental
       importance, a substantial effort should be devoted to attacking this
       problem.

\item[Algorithmic improvements]
      All these developments recommended above depend on the
      efficient and robust generation of self-consistent mean-field solutions with
      many constraints. Constructing large numbers of these configurations is
      extremely demanding both in CPU time and in the time required to diagnose
      convergence issues.  Besides the second-order gradient method
      \cite{(Rob11c)}, advanced algorithms from the field of non-linear
      optimisation have been introduced to accelerate the convergence of
      self-consistent iterations \cite{(Bar08),Ryssens2019}; the
      potential for further developments could be far greater. Supervised or
      deep learning techniques could also present significant opportunities to,
      e.g., optimise basis parameters for better numerical accuracy, perform
      real-time diagnostic about convergence, or provide good emulators of
      theoretical models \cite{(Las20)}.

\item[High-Performance Computing]
      Already in the late 2000s, nuclear fission was recognised as a scientific grand
      challenge justifying the development of exascale computer systems
      \cite{young2009,bishop2009,exascale2016}. Many of the various recommendations discussed
      in this document, from large-scale potential energy surfaces with many
      degrees of freedom, to the coupling between TDHFB and TDGCM dynamics, will
      require the power of such facilities. Yet, this will require a serious
      effort by the community to modify, or re-factor, their codes in order to
      adapt to choices made at leadership computing facilities. Such choices
      include hardware architectures (GPU and hybrid chips, memory/core),
      software libraries (use of abstraction layers, more and more often in
      C++), or computing policies (limited runtime).
\end{description}

In conclusion, we recommend the development of numerical tools, which
(i) target specific requirements of fission dynamics within the single-reference and multi-reference
frameworks; (ii) are adapted to modern computing infrastructure; and
(iii) build a common code base for fission theory. We advocate to
increase the transparency associated with numerical choices by: (i)
including a detailed description of numerical procedures in published studies
(for benchmarking and an independent reproduction of the results);
(ii) making codes publicly available under Open Source license along with their long-term
continuous maintenance within, e.g.,  a git
repository; and (iii) writing the codes in a sufficiently modular fashion, so that new
advances can be more easily adopted by the community.


\subsubsection{Databases}
\label{subsec:databases}

As discussed in {\Sec}~\ref{subsubsec:new}, the computational cost to
generate and store the mean-field configurations and their related
quantities can be extremely high in the context of fission
applications. At the same time, many applications only require
``integral'' quantities related to these configurations. For example,
computing spontaneous fission lifetimes in the WKB approximation, or
fission fragment distributions within the TDGCM+GOA framework, only
requires the HFB PES and the collective inertia
tensor. Let us assume for the sake of the argument a 3-dimensional
collective space with 200$\times$50$\times$50 = 500 000 points.
Lifetimes, barrier penetrabilities, charge and mass distributions of
the fission fragments can be computed merely from the knowledge of 1
scalar function of 3 variables (the energy) and a rank-2 tensor
function of 3 variables (the inertia tensor). In our example, we
would only need a grand total of 3.5$\times 10^{6}$ function values,
which takes a very small amount of storage. By contrast, storing all
the information about the HFB solution across the same PES would take
up in excess of 1.6$\times 10^{13}$ function values (assuming a 40$\times$20$\times$20 box
discretisation of 2000 HFB spinors). Having a database of such
potential energy surfaces for nuclei, different energy functionals
(Skyrme, Gogny, relativistic, different pairing functionals, etc.) and
different DFT solver technologies would be very valuable to quantify
theoretical uncertainties. Since the cost of generating a PES is
high, it would also offer maximum leverage to our small community.


\section{Recommendations and challenges}
\label{sec:summary}

The purpose of this section is to summarise the main recommendations of this report
that reflect challenges facing nuclear fission theory.
The high-level recommendations, addressing the general challenges
facing the field of microscopic nuclear fission theory,  are listed
in {\Sec}~\ref{highlevel}. More detailed recommendations, pertaining
to specific subareas, are listed in {\Sec}~\ref{specific}. The
ordering does not imply any priority.

\subsection{High-level recommendations}\label{highlevel}

\noindent
General  recommendations relevant to the field as a whole follow below. The numbers in brackets refer to key sections pertaining to individual recommendations.

\begin{description}

\item[Quantified input]
Quantitative predictions require quantified input. It is essential to develop
interactions and  energy
density functionals  that are
specifically tailored for the purpose of modelling fission.
Of particular importance are the interaction components responsible for nuclear deformability and the pairing interactions that control the level of adiabaticity.
 It would
be desirable to develop several quantified interactions/functionals for fission
studies for (i) benchmarking purposes and (ii) to assess statistical
and systematic uncertainties. Moreover, statistical  calibration of interactions for fission should be carried out
that would determine the sensitivity of parameters to key experimental constraints.
[\ref{sec:EDF}, \ref{sec:inputs}]

\item[Focus on essential ingredients]
Considering limited resources, in order to maximise  progress it is important to identify the essential ingredients in fission theory that require
careful microscopic treatment and   more robust ingredients that are
necessary for a correct description of fission dynamics, but perhaps  require
less sophisticated modelling at the early stage of development.
An example of essential ingredient is the PES. A more robust quantity is dissipation tensor; indeed many properties of predicted fission yields
are found insensitive to large variations of the dissipation strength.
[\ref{sec:inputs}]

\item[Modern theory extensions]
There exist  microscopic, yet greatly unutilised, extensions  of  current models of non-adiabatic large-amplitude collective
motion  that can be adopted to modern studies of nuclear fission. Many of those techniques involve  algorithmic developments and significant computational capabilities.
This includes: (i) Description of
fission trajectories in the full TDHFB manifold; (ii) Inclusion of
non-adiabatic couplings between many-body configurations; and (iii)
Consistent treatment of quantum  and statistical fluctuations.
[\ref{sec:concepts}, \ref{sec:inputs}, \ref{sec:entrance}, \ref{sec:computation}]

\item[Comprehensive description]
Fission is a complex  phenomenon with a multitude of final channels
and measured observables. In order not to be misled by a good
agreement with limited classes of data, it is advisable to develop a
comprehensive approach to fission observables.  This is important
because different elements of fission models are sensitive to
different data. For instance, good reproduction of fission yields
does not guarantee quality predictions of TKEs. In this context,
priority should be given to modelling of measured quantities, not
unobservables, which are primarily of theoretical interest.
[\ref{sec:fragments}, \ref{subsec:observables}]

\item[Access to quantum numbers]
To be able to describe fission observables, a connection between models of fission dynamics based on the
intrinsic-system concept, and the symmetry-conserved observables studied
experimentally (particle number, angular momentum, parity) needs to be established.
There are two possible avenues to achieve this goal. One is based on
a reaction-theory approach that is explicitly formulated in the
laboratory reference frame. Another way is by means of projection
techniques. In both cases, many foundational developments are needed.
[\ref{sec:concepts},\ref{sec:fragments}]

\item[Entrance channels]
To model various kinds of fission, it is important
to develop a unified description of initial states.  At various instances  of the fission phenomenon (from  spontaneous fission to  fission
induced by fast probes; from low-energy to high-energy fission), the entrance channel should be properly described. This includes the realistic modelling of compound nucleus for neutron-induced fission
as well as specific nuclear states populated in photofission or $\beta$-decay. In the latter case,
implementation of flexible  QRPA methods (for any shape,
for arbitrary multipole and charge-exchange
channels, and indiscriminately for even-even, odd, and odd-odd
systems) is recommended.
[\ref{sec:entrance}]

\item[Computing]
Future exascale computing ecosystems will offer  a unique opportunity for
microscopic modelling of nuclear fission.  To achieve this goal, this report
recommends
the development of  specific computing capabilities and launching
a library of general-purpose fission software based on novel algorithms  and
programming that can efficiently utilise modern computing
infrastructures. To this end, collaborations with computer scientists,
 applied mathematicians, and data scientists  will be needed to (i)  develop
open-source, modular nuclear solvers and (ii) leverage
high-performance computing and  statistical machine learning.
[\ref{subsec:codes}]

\item[Databases]
Establish  databases of microscopic fission output for further
processing. This can include various HFB and TDHFB results (PESs,
fission pathways, fission fragment yields and properties).
Having computed multi-model fission data available
will be essential not only for post-processing but also for benchmarking and uncertainty quantification.
[\ref{subsec:databases}]

\end{description}

\subsection{Specific recommendations}\label{specific}

\noindent
A number of specific recommendations are proposed in the
body of this report. The numbers in brackets refer to specific sections where individual recommendations can be found.

\begin{description}

\item[Microscopic tunnelling]
In the studies of SF and  low-energy fission, a  part of collective
motion proceeds through  the classically-forbidden space.  The most
commonly used approach is that based on the CSE  and a WKB
approximation, in which the tunnelling rate is obtained from the
collective action calculated along an effective one-dimensional
trajectory. Limitations of this approach should be studied and,
depending on the outcome, extensions explored. Those
include: generalisation of  the one-dimensional  WKB treatment to
several dimensions; increasing the number of collective coordinates;
or other approaches to tunnelling, such as the imaginary-time method.
[\ref{subsec:tunnelling}]

\item[Classical aspects of TDDFT]
Since TDHF equations emerge as a classical field theory for interacting
single-particle fields, the TDDFT approach can neither  describe the
motion of the system in classically-forbidden regions  of the collective
space nor quantum fluctuations. In the context of tunnelling, one
should determine the feasibility of arriving at instanton solutions
to the TDDFT fission problem and develop methods to calculate the
full ATDHFB collective inertia. As far as fluctuations are concerned,
this problem shows up in too-narrow  fission yield distributions predicted  by
time-dependent theories. A  possible resolution to this problem lies
in the Stochastic Mean-Field approach  that allows larger
fluctuations in collective space. [\ref{sec:concepts}]

\item[Extend theory beyond even-even systems]
Most microscopic calculations of nuclear fission pertain to even-even
nuclei. It is therefore urgent to develop a consistent theoretical
framework for the fission of even-even, $A$-odd, and odd-odd nuclei.
This will require going beyond the  usual blocking  approximation to
fully consider  time-reversal symmetry-breaking effects. Odd-even
staggering of fission yields is an example of a quantity that can be
sensitive to  such effects. [\ref{sec:concepts}, \ref{sec:fragments}]

\item[Microscopic Langevin approach]
Classical Langevin theory has been very successful in explaining many
properties of fission products. To bridge it with microscopic fission
frameworks, it is important to clarify the connections between
microscopic TDHFB and TDGCM with dissipative theories -- to make
contact with Langevin-based  approaches. [\ref{subsec:dissipation},
\ref{sec:fragments}].

\item[Reaction-theory framework]
An approach to fission based on reaction theory is useful, because it is
explicitly formulated in the laboratory reference frame, which
guarantees that the important quantum numbers are conserved. One
should consider assessing the feasibility of developing a practical
microscopic approach to fission based on the $K$-matrix reaction
theory.  It offers a completely different calculational framework for
spontaneous fission as well. [\ref{subsec:reaction}]

\item[Generalised fission path]
On the way from the entrance configuration to scission, the
fissioning nucleus explores the continuum of trajectories in the
collective space. Current approaches explore limited sectors of this
space and hence it is essential to develop methods to search for
optimum fission pathways in such a way that a blind exploration of
the full multi-dimensional collective space is not required.
[\ref{subsec:dof}]

\item[Generalised constraints]
It would be very useful to go beyond simple constraining operators for which
important configurations may be overlooked.  Within
the large family of density constraints, the technique
that constrains the entire density distribution obtained in TDDFT
is promising in that it
provides a tight control of the shape.  It naturally localises
the system in the space of nuclear configurations, as does
wave function constraints such as the $K$-partitioning.  Also,
constraints based on fission observables may be useful in the
study of fluctuations.
[\ref{subsubsec:density-constrained}]

\item[Residual interactions]
The ability to compute Hamiltonian
matrix elements between configurations is essential for microscopic
calculations of reaction theory, level
crossing dynamics, and the dissipation tensor.  This capability is already included
for the pairing interaction in CHFB.  However, the neutron-proton
interaction is ignored in current codes except for its mean-field
contribution.
[\ref{subsec:reaction}, \ref{Landau-Zener}, \ref{subsubsec:friction}]

\end{description}

\section{Summary and Conclusions}
\label{sec:summary2}

This Topical Review is unconventional: rather than  presenting past
achievements, it aims at reviewing future options for theory of
nuclear fission. As we discussed in the Introduction, numerous
reviews and books that discuss experimental and theoretical aspects
of nuclear fission  exist
\cite{(Sch16b),Bertsch2015,(And17),Talou2018,Schmidt2018,krappe2012,younes2019}.
The interested reader is encouraged to consult these references for
more extensive and detailed information. The goal of the present work
was to lay down our opinions on directions of future research in
theory of nuclear fission. While this task is challenging, we found
it useful to talk about research directions that seem to be promising
and at the same time may be long overdue.

We will be most happy if our ideas are picked up by enthusiastic
researchers working in this domain of nuclear physics, and even more,
if they attract new talent into this area. Undoubtedly, many proposed
directions will require concerted efforts of large collaborations,
and we hope that this Topical Review will contribute to fostering
those.

Beyond phenomenological modelling, the theoretical description of
nuclear fission requires novel ideas on how to treat the incredible
complexity of the phenomenon in a manageable and physically
meaningful way. To be implemented, many of those ideas require
advanced computations. In this document, we call for performing
baseline work on developing quantified interactions/functionals for
fission studies; identifying the essential ingredients in fission
theory; utilizing extensions  of current models of non-adiabatic
large-amplitude collective motion; developing a comprehensive
approach to fission observables; making connection between models
based on the intrinsic-system concept and the symmetry-conserved
observables; and for realistic modelling of the compound nucleus as
well as of the specific nuclear states that form gateways to fission.
Each and every one of these projects may in the future become a
challenging research direction. Together, they may lead to a major
advance of the field.

\section{Acknowledgements}

ASU and CS  would like to thank Prof.\ David Hinde for useful discussions regarding the neutron clock.
This work was partially supported
by the STFC grant Nos.~ST/M006433/1, ST/P003885/1,
and ST/P005314/1,
by the Polish National Science Center under Contract Nos.~2018/31/B/ST2/02220,
2018/30/Q/ST2/00185,
and 2017/27/B/ST2/02792;
by JSPS KAKENHI Grant Number JP19K03861;
by the National Natural Science Foundation of China under grant Nos.\ 11875225 and 11790325;
by the U.S. Department of Energy under grant Nos.~DE-SC0013847,
DE-SC0019521, DE-SC0013365, DE-SC0018083, DE-NA0003885,
and DE-SC0019521;
by Spanish Ministry of Economy and Competitiveness (MINECO)
grant No.~PGC2018-094583-B-I00;
by the Fonds de la Recherche Scientifique (F.R.S.-FNRS)
and the Fonds Wetenschappelijk Onderzoek - Vlaanderen (FWO)
under the EOS Project nr O022818F;
and by the Australian Research Council grant No.~DP190100256.
This work was partly performed under the auspices of the U.S.\
Department of Energy by Lawrence Livermore National Laboratory under Contract
DE-AC52-07NA27344 (NS).
The work was also supported by the US Department of Energy
through the Los Alamos National Laboratory.  Los Alamos National Laboratory
is operated by Triad National Security, LLC, for the National Nuclear
Security Administration of U.S.\  Department of Energy (Contract No.\
89233218CNA000001).
This project has received funding from the European Union's Horizon 2020 research and
innovation programme under Grant Agreement No.\ 654002.
JR wishes to acknowledge the Yukawa Institute for Theoretical Physics
in Kyoto for its generous support of his participation is this project.

\newpage

\bibliographystyle{jphysicsB-withTitles}

\end{document}